\documentclass[twocolumn,showpacs,preprintnumbers,
prb,groupedaddress,nofootinbib]{revtex4-2}%
\usepackage{eucal}
\usepackage{amssymb}
 \usepackage{relsize}
\usepackage{dsfont}
\usepackage{mathtools} 
\usepackage{stfloats}
\usepackage{calligra,amsmath,amssymb,amsthm}
\usepackage{graphicx}
\usepackage{float}
\usepackage{dcolumn} 
\usepackage{mathbbol}
\usepackage{feynmf}
\usepackage{bm}
\usepackage{scalerel}
\usepackage{bbm}
\usepackage[makeroom]{cancel}
\usepackage{mathrsfs} 
\usepackage[export]{adjustbox}
\usepackage{amsfonts}%
\usepackage{lipsum}
\usepackage[usenames, dvipsnames]{color}
\usepackage{chngcntr}
\usepackage[T2A,T1]{fontenc}
\usepackage[utf8]{inputenc}

\newcommand{\smalltimes}[1][0.8]{%
    \mathbin{\vcenter{\hbox{\scalebox{#1}{$\times$}}}}}

\counterwithout{equation}{section}
\providecommand{\U}[1]{\protect\rule{.1in}{.1in}}

\newcommand{\bcdot}{\boldsymbol{\cdot} }

\newcommand{\half}{\scalebox{1}{ \text{$\frac{1}{2}$}}}

\newcommand{\smp}{\scalebox{0.5}{\text{$\! + \! \!$}}}

\newcommand{\lp}{\scalebox{1}{\text{$^{\!}(^{\!}$}}}
\newcommand{\rp}{\scalebox{1}{\text{$^{\!})$}}}

\newcommand\varpm{\mathbin{\vcenter{\hbox{%
  \oalign{\hfil$\scriptstyle+$\hfil\cr
          \noalign{\kern-.3ex}
          $\scriptscriptstyle({-})$\cr}%
}}}}

\numberwithin{equation}{section}

\begin{document}

\setlength{\belowdisplayskip}{12pt} 
\setlength{\belowdisplayshortskip}{0pt}
\setlength{\abovedisplayskip}{12pt} 
\setlength{\abovedisplayshortskip}{0pt}


\title{Magnon-phonon coupling from a crossing symmetric screened interaction}

\author{T. J. Sj\"{o}strand}
\affiliation{Department of Physics, Division of Mathematical Physics, Lund University,
Professorsgatan 1, 223 62 Lund, Sweden}
\author{F. Aryasetiawan}
\affiliation{Department of Physics, Division of Mathematical Physics, Lund University,
Professorsgatan 1, 223 62 Lund, Sweden}
\date{\today }

\begin{abstract}
The magnon-phonon coupling has received growing attention in recent years due to its central role in spin caloritronics and the emerging field of acoustic spintronics. At resonance, this magnetoelastic interaction drives the formation of magnon polarons, which underpin exotic phenomena such as magnonic heat currents and phononic spin, but has with a few recent exceptions only been investigated using mesoscopic spin-lattice models. Motivated to integrate the magnon-phonon coupling into first-principle many-body electronic structure theory, we set up to derive the non-relativistic exchange-contribution, which is more subtle than the spin-orbit contribution, using Schwinger's method of functional derivatives. To avoid having to solve the famous Hedin-Baym equations self-consistently, the phonons are treated as a perturbation to the electronic structure. A formalism is developed around the idea of imposing crossing symmetry on the interaction, in order to treat charge and spin on equal footing. By an iterative scheme, we find that the spin-flip component of the {\it collective} four-point interaction, $\mathcal{V}$, which is used to calculate the magnon spectrum, contains a first-order ``screened T matrix'' part and an arguably more important second-order part, which in the limit of local spins describes the same processes of phonon emission and absorption as obtained from phenomenological magnetoelastic models. Here, the ``order'' refers to the {\it screened collective} four-point interaction, $\mathcal{W}$ --- the crossing-symmetric analog of Hedin's $W$. Proof-of-principle model calculations are performed at varying temperatures for the isotropic magnon spectrum in three dimensions in the presence of a flat optical phonon branch.
\end{abstract}

\pacs{71.2 0.-b, 71.27.+a}
\maketitle

\section{Introduction}
Magnons and phonons are ubiquitous bosonic quasiparticles in condensed matter physics. Magnons denote collective spin flips --- spin fluctuations, with angular momentum $\hbar$ and with dispersion and life-time accessed by the magnetic susceptibility, commonly probed by inelastic neutron scattering. Phonons denote collective lattice deformations, carry no angular momentum in the absence of spin-phonon interaction and are accessed by the charge susceptibility, also commonly probed by inelastic neutron scattering. The two quasiparticles have very distinct physical properties and applications. 

Phonons are first and foremost major heat carriers and prevent overheating in microelectronics, although a major issue is that downscaled electronics has a large electron-phonon coupling that inhibits the heat dissipation. The electron-phonon coupling also has positive usages, such as the boosting of thermoelectric generators to reduce environmental waste \cite{liao2016}. Phonons are employed in cavity optomechanics, with applications ranging from gravitational wave detection to quantum metrology, where phononic crystal cavities are often used to confine sound \cite{Eichenfield2009}, and laser detuning is used to transfer energy to or from cavity phonons for mechanical amplification or cooling purposes \cite{aspel}. Magnons, on the other hand, are well known to dissipate Zeeman energy and relax magnetization, but are also used practically for information transport and processing in the emerging field of magnonics, which aims to achieve downscaled and faster computing by pushing for increased magnon speeds, life times and mean-free paths \cite{serga2010,cornelissen2015}. A difficulty in magnon spintronics/electronics is converting information stored in electronic spin/charge to and from the processing magnonic subsystem \cite{chumak2015}. Magnons also allow for room-temperature condensation when exposed to microwave pumping \cite{demokritov2006,ruckrieger2015}, and, through their scattering with electrons, magnons work as a possible pairing glue in unconventional superconductors \cite{scalapino2012,essenberger2014}. It so follows that a unified picture of superconductivity necessarily be equipped with a proper account of both phonons and magnons. The two can generally not be adiabatically separated, which correlates their dynamics and opens the door to exotic phenomena with novel applications. 

The magnon-phonon coupling is strong in manganites \cite{dai2000}, nickel nanomagnets and nanopatterned magnetic structures \cite{berk2019,berk2020}, yttrium iron garnet \cite{man2017}, polycrystalline BiFeO$_3$ \cite{ramachandran2015},
Eu$_{0.75}$Y$_{0.25}$MnO$_3$ \cite{aguilar2007} and many other multiferroics, but also in ferromagnets such as bcc iron \cite{Kormann2014,Perera2017}. Depending on the relative scattering strength of magnons and phonons in magnetic insulators, the coupling can either increase or decrease the spin Seebeck effect \cite{jaworski2011,kikkawa2016,flebus2017,yahiro2020,uchida2011}, a thermoelectric effect that converts temperature gradients to spin currents. This puts the magnon-phonon coupling at the core of spin caloritronics \cite{bauer2012}, where coupled spin-heat currents are studied --- a path to green devices with waste heat recycling capabilities. Another promising direction is acoustic spintronics, initiated by the room-temperature ``spintronics battery'' in yttrium iron garnet, based on acoustic spin pumping by magnons-phonon resonance \cite{weiler2012,hayashi2018}. The magnon-phonon coupling also leads to thermal conductivity increase upon magnetic ordering in geometrically frustrated magnets \cite{sharma2004}, magnonic heat currents \cite{katsura2010} and phononic spin \cite{zhang2014,Garanin2015,holanda2018} in quantum magnets, and enables photodrive of curved domain walls \cite{ogawa2015}. It has also provided a mechanism for thermal Hall effect \cite{zhang2019}, where local non-equilibrium between magnons and phonons can be achieved optically \cite{an2016}. We conclude this acclamation of the magnon-phonon coupling by mentioning its link to the coexistence of superconductivity and charge-density waves in high-temperature superconducting copper oxides \cite{struzhkin2016} and to condensation of hybrid magnetoelastic bosons \cite{bozhko2017}.

Prior to such exciting developments, the magnon-phonon coupling was mainly theoretically investigated using spin-lattice models, owing to the success of the Landau-Lifshitz-Gilbert approach to describe mesoscopic magnetization dynamics \cite{lifshitz1935}, where precession and damping are included but faster superposed effects are neglected. The initial field-theoretic works in this direction by Abrahams and Kittel were focused on ferromagnets \cite{abrahams1952,abrahams1954,kittel1958}, and based on tuning a macroscopic magnon-phonon coupling obtained from a postulated magnetoelastic free-energy density to match observed magnetostriction, and of a pseudo-dipolar spin-spin interaction to match observed anisotropy. The famous phonon emission and absorption terms were identified as leading terms. Since then, many spin-lattice model studies have contributed to further insights into the coupling in ferro-, ferri-, antiferro- and metamagnets, as well as in complicated set-ups, in and out of equilibrium \cite{kaganov1959,tiersten1964,schlomann1964,pytte1965,silberglitt1969,rezende1969,guerreiro1971,jensen1975,economou1975,sanger1994,woods2001,cheng2007,berciu2009,ruckriegel2014,Guerreiro2015,kamra2015,chernyshev2015,takahashi2016,flebus2017,Maehrlein2018,schmidt2018,holm2018,rameshti2019,Streib2019}. A general finding is that the exchange-mediated contribution tends to dominate in the short-wavelength limit. An important step towards first-principle descriptions of spin-lattice dynamics was taken within time-dependent density functional theory \cite{antropov1996}, where molecular dynamics was studied in conjunction with an adiabatically separated spin-density matrix, albeit with a limited treatment of dissipation and temperature. Another important step was the extension of the Landau-Lifshitz-Gilbert approach to account for short-time nutation in the magnetization dynamics \cite{Ciornei2011}, caused by moment of inertia and necessitating microscopic treatments of the exchange interaction where retardation is included \cite{Bhattacharjee2012}. 

In this work, we focus on the purely exchange-mediated magnon number-conserving magnon-phonon coupling, and therefore neglect spin-orbit coupling and the associated anisotropy. We focus on the magnon spectrum, and treat the phonons as perturbations to avoid having to self-consistently solve the full Hedin-Baym equations \cite{Giustino2017} for coupled electron and phonon systems. This allows for the phonons to be accounted for at the end of the derivation, by a method of replacements \cite{hedin69}. A manifestly crossing-symmetric formalism based on that of Hedin is presented, where charge and spin are treated on equal footing. We also go through the approximations needed to arrive at a coupling of the same form as obtained from phenomenological magnetoelastic models. The most crucial step is the two-point approximation, which reduces the spin-flip interaction from a four-point to a two-point quantity. By applying the theory to a simple model, we show how the purely exchange-mediated magnon-phonon coupling not only leads to dissipative broadening of the magnon spectrum but also to splittings. In an upcoming publication, a semirelativistic extension will be presented where the orbital magnetic moment and the spin-orbit coupling will be included in addition to the exchange-mediated coupling considered in this work.

This paper is organized as follows: in Sec. II we present the formal theory based on Schwinger functional derivatives, where crossing symmetry is used to put charge and spin on equal footing. In Sec. III a series of approximation are made, which allow for a simple expression of the magnon number-conserving magnon-phonon coupling. In Sec. IV, a minimal model is considered in which the effect of phonons on the magnon spectrum is considered. Finally, in Sec. V we summarize the work and give an outlook on future developments. 
\counterwithout{equation}{section}
\section{Formal theory}
\subsection{Setting the scene}\label{secc0}
The main goal of this work is to derive the magnon-phonon coupling from first principles in the absence of spin-orbit coupling or other relativistic effects. We assume that the orbital angular momentum is quenched in the solid, and therefore assume that the magnons are exclusively attributed to the spin-fluctuations. The non-relativistic contribution to the coupling originates from the exchange interaction, and is obtained self-consistently from the Hedin-Baym equations for systems where both the electrons and the lattice fluctuate \cite{Giustino2017}. Unfortunately, this self-consistency not only makes practical calculations costly, but also clouds the connection to spin-lattice models.

However, it is well-known that due to the high-energy (plasmonic) charge screening of the electronic subsystem it is crucial to calculate the phonon Green's function $D$ after an initial electronic structure calculation. Having obtained phonons in such a simplified electronic surrounding it is natural to compute the electronic structure in the presence of these approximate phonons. This can easily be done in Hedin's formalism by correcting the well-known screened electron-electron interaction $W$ by an additional term $WDW$, containing an intermediate phonon \cite{hedin69}. Since the effects of the approximate phonons exclusively enter through this quantity it is possible to first derive the spin-fluctuation spectrum as a functional of $W$, i.e. in a rigid lattice, and then replace the latter by $W+WDW$ to account for the magnon-phonon coupling. We can thus forget about the lattice dynamics completely for now --- the effect from it is easily corrected for at the end. In principle, it is possible to compute how the resulting magnons affect both the electronic structure as well as the ``next iteration phonons'', but in this work we focus on the effect of phonons on magnons. 
\subsection{A crossing symmetric starting point}\label{secc}
Since equilibrium magnons and phonons are excited thermally, it is natural to work with a finite-temperature formalism, where an appropriate starting point is the grand canonical Hamiltonian, defined as $\hat{K}= \hat{H}- \mu \hat{N}$, where $\hat{H}$ is the Hamiltonian of the electronic system, $\mu$ the chemical potential and $\hat{N}$ the number operator. In an orthonormal Wannier basis, this takes the form  
\begin{align} \label{eqnew1}
\hat{K}  & =   \big( h  (  \textbf{\textit{12}} )  -  \delta_{ \textbf{\textit{12}}}^{} \mu   \big) \hat c_{\sigma_\mathit{1}}^{ \dag} ( \textbf{\textit{1}} ) \hat c_{\sigma_\mathit{1}}^{} ( \textbf{\textit{2}} )  
\\ & + \half v(\textbf{\textit{12}} , \! \textbf{\textit{34}})  \hat 
c_{\sigma_\mathit{1}}^{\dag}(\textbf{\textit{1}}) \hat c_{ \sigma_\mathit{4}}^{\dag}(\textbf{\textit{4}}) \hat c_{ \sigma_\mathit{4}}^{}(\textbf{\textit{3}}) \hat c_{\sigma_\mathit{1}}^{}(\textbf{\textit{2}})   , \nonumber
\end{align}
where $h$ is the hopping matrix and 
\begin{align} \label{eqtwo}
  v \lp \textbf{\textit{12}},\! \textbf{\textit{34}} \rp  &  =   \!  \int  \! d{\boldsymbol x} d{\boldsymbol y} \frac{w_{n_\mathit{1 } {\boldsymbol R}_\mathit{1 \!} }^{*}  \lp {\boldsymbol x} \rp w_{n_\mathit{2  }  {\boldsymbol R}_\mathit{2} \! }^{}  \lp {\boldsymbol x} \rp  w_{n_\mathit{3} {\boldsymbol R}_\mathit{3 \!} }^{}  \lp {\boldsymbol y} \rp w_{n_\mathit{4 } {\boldsymbol R}_\mathit{4 \!}  }^{*}  \lp {\boldsymbol y} \rp}{|{\boldsymbol x}-{\boldsymbol y}|}  
\end{align}
\\[-0.3cm]
the Coulomb integrals in the Wannier basis $\{ w_{n_i {\boldsymbol R}_i } \}$, which fulfil the inversion symmetry $v(\textbf{\textit{12}},\textbf{\textit{34}}) = v(\textbf{\textit{43}},\textbf{\textit{21}}) $. The Wannier functions are assumed spin-independent for simplicity. In Eq. \eqref{eqnew1} and the following, we use Einstein's summation convention and Hartree's atomic units, where $\hbar=m_e=e=a_0=1$, and let the electron charge be $+1$ rather than $-1$, for convenience. The spin index $\sigma_i=\uparrow,\downarrow$ is kept explicit whereas the Wannier orbital $n_i$ and unit cell ${\boldsymbol T}_i$ are condensed into index $\textbf{\textit{i}}$. To describe the interplay between charge and spin degrees of freedom it is key to start from a crossing symmetric interaction which encodes the Pauli principle in its definition. This utilizes that the local part of a spin-conserving interaction between a spin $\downarrow$ and $\uparrow$ electron can be reinterpreted as a simultaneous spin-flip of the two --- both processes conserving the total spin locally. 

Using anticommutation, it is possible to reformulate the interaction term in Eq. \eqref{eqnew1}, which we call $\hat{K}_{int}$, into a crossing symmetric form. The result is  
\begin{align}   \label{eqKint}
\hat{K}_{int}  & =  \half v^{\sigma_\mathit{1} \sigma_\mathit{2}}_{\sigma_\mathit{3} \sigma_\mathit{4}} (\textbf{\textit{12}},\textbf{\textit{34}})  \hat 
c_{\sigma_\mathit{1}}^{\dag}(\textbf{\textit{1}}) \hat c_{ \sigma_\mathit{4}}^{\dag}(\textbf{\textit{4}}) \hat c_{ \sigma_\mathit{3}}^{}(\textbf{\textit{3}}) \hat c_{\sigma_\mathit{2}}^{}(\textbf{\textit{2}})  .  
\end{align}
The spin-dependent interaction can be written on the Pauli form ($\mu=0,x,y,z$)  
\begin{align}  \label{crossing}
v^{\sigma_\mathit{1} \sigma_\mathit{2}}_{\sigma_\mathit{3} \sigma_\mathit{4}} (\textbf{\textit{12}},\textbf{\textit{34}}) & = - v^{\sigma_\mathit{1} \sigma_\mathit{3}}_{\sigma_\mathit{2} \sigma_\mathit{4}} (\textbf{\textit{13}},\textbf{\textit{24}}) \\ 
\label{eqvnajs} 
 & = \sigma^{\mu_1}_{\sigma_\mathit{1} \sigma_\mathit{2}} v_{\mu_1 \mu_\mathit{2}}^{}(\textbf{\textit{12}},\textbf{\textit{34}}) \sigma^{\mu_\mathit{2}}_{\sigma_\mathit{4} \sigma_\mathit{3}} ,  
\end{align}
where the first equality shows the crossing symmetry, and 
\begin{align}
 v_{\mu_\mathit{1} \mu_\mathit{2}}^{} \lp \textbf{\textit{12}},\textbf{\textit{34}} \rp & ^{\!} = ^{\!} \frac{  \delta_{\mu_\mathit{1} \mu_\mathit{2}}}{2} ^{\!}\big(   \delta_{\mu_\mathit{1} 0}^{}    v \lp \textbf{\textit{12}},\textbf{\textit{34}} \rp   \! - \!  \tfrac{1}{2}    v \lp \textbf{\textit{13}},\textbf{\textit{24}} \rp   \big)      .
\end{align} 
This is the origin of the exchange interaction between spins in the Heisenberg model. It forbids unphysical local interactions between two electrons of identical spin in the many-body treatment, which is to follow, and is thus the most appealing form allowed by the Fierz ambiguity \cite{fierz}.
\subsection{Schwinger's functional derivative method}\label{secc2}
We now use Schwinger functional derivatives, with the goal of obtaining the magnon spectrum from the electronic structure. The starting point is the imaginary-time Green's function for an electron in the Dirac picture 
\begin{align} \label{eqG}
\CMcal{G}_{\sigma_\mathit{1} \sigma_\mathit{2}}^{}(\mathit{12})  &  = - \frac{  \operatorname{Tr} \big( e^{-\beta  \hat{K} } \hat{\CMcal{T}}   \hat{\CMcal{S}}   \hat c_{\sigma_\mathit{1}}^{}(\mathit{1})     \hat c_{\sigma_\mathit{2}}^\dag(\mathit{2})    \big)   }{
\operatorname{Tr} \big( e^{-\beta  \hat{K}}  \hat{ \CMcal{S} }    \big)
} ,
\end{align}
where the imaginary times $\tau_\mathit{i}$, which are hidden in the combined indices $\mathit{i}=(\textbf{\textit{i}}, \tau_\mathit{i})$, are assumed to be between $0$ and the ``thermodynamic beta'', $\beta=1/k_B T$. Furthermore, $\hat{\CMcal{T}}$ is the time-ordering operator in imaginary time, $\hat{c}_{\sigma_\mathit{1}}^{}(\mathit{1})  = e^{\hat{K} \tau_\mathit{1}} \hat{c}_{\sigma_\mathit{1}}^{}(\textbf{\textit{1}}) e^{- \hat{K} \tau_\mathit{1}} $ the annihilation operator in the Dirac picture, and $\hat{\CMcal{S}}$ the imaginary-time evolution operator from 0 to $\beta$, due to a virtual (external) two-point field $\varphi^{ext}$, i.e.
\begin{align} \label{eqSa}
{\hat{\CMcal S}} & = \hat{ \CMcal{T}}  \exp \big(   - \varphi_{\sigma_\mathit{3} \sigma_\mathit{4}}^{ext}(\mathit{34}) \hat c_{\sigma_\mathit{3}}^{ \dag}(\mathit{3}) \hat c_{\sigma_\mathit{4}}^{}(\mathit{4})  \big) ,
\end{align}
with implicit integration from $0$ to $\beta$ over both $\tau_\mathit{3}$ and $\tau_\mathit{4}$. The equation of motion for $\CMcal{G}$ can easily be obtained from the Heisenberg equation, which reads 
\begin{align} \label{heis}
- \partial_{\tau_\mathit{1}} \hat{c}_{\sigma_\mathit{1}}(\mathit{1}) & = \big[  \hat{c}_{\sigma_\mathit{1}}(\mathit{1}) , \hat{K} \big] .
\end{align}
This commutator is evaluated using Eqs. \eqref{eqnew1} and \eqref{eqKint}, and results in the equation of motion 
\begin{align} \label{eqomm}
  & \hspace{3.3cm} \delta_{\sigma_\mathit{1} \sigma_\mathit{2}}  \delta_{\mathit{12}} = \\
    &   -  \Big(\delta_{\sigma_\mathit{1} \sigma_\mathit{3}}^{} \big(   \delta_{\mathit{13}}^{}\partial_{t_\mathit{1}}^{} + k (\mathit{13})
 \big)    + \varphi_{\sigma_\mathit{1} \sigma_\mathit{3}}^{ext}(\mathit{13}) \Big) \CMcal{G}_{\sigma_\mathit{3} \sigma_\mathit{2}}^{}(\mathit{32})  \nonumber    
\\
 + &   v^{\sigma_\mathit{1} \sigma_\mathit{3}}_{\sigma_\mathit{4} \sigma_\mathit{5}}
(\mathit{13},\mathit{45}) \frac{  \operatorname{Tr} \big( e^{-\beta \hat{K} }  \hat{\CMcal{T}}   \hat{\CMcal{S}}   \hat c_{ \sigma_\mathit{5}}^{ \dag}(\mathit{5}^{\smp})  \hat c_{ \sigma_\mathit{4}}^{}(\mathit{4}) \hat c_{\sigma_\mathit{3} }^{}(\mathit{3}) \hat c_{\sigma_\mathit{2}}^{ \dag}(\mathit{2})  \big)  }{ \operatorname{Tr} \big( e^{-\beta \hat{K} }   \hat{\CMcal{S}}   \big)}  \nonumber ,
\end{align}   
where we have defined ($\delta_{\tau_\mathit{1} \tau_\mathit{2}}=\delta(\tau_\mathit{1} - \tau_{\mathit{2}})$)
\begin{align}
k (\mathit{13}) & = \delta_{\tau_\mathit{1} \tau_\mathit{3}}^{}  \big( h (\textbf{\textit{13}}) -  \delta_{ \textbf{\textit{13}}}^{} \mu   \big)  , \\
\label{eqv0}
v^{\sigma_\mathit{1} \sigma_\mathit{3}}_{\sigma_\mathit{4} \sigma_\mathit{5}}
(\mathit{13},\mathit{45}) & = \delta_{\tau_\mathit{1} \tau_\mathit{3}}^{} \delta_{\tau_\mathit{3}  \tau_\mathit{4}}^{} \delta_{\tau_\mathit{4} \tau_\mathit{5}}^{}  v^{\sigma_\mathit{1} \sigma_\mathit{3}}_{\sigma_\mathit{4} \sigma_\mathit{5}} (\textbf{\textit{13}},\textbf{\textit{45}}) .
\end{align}
The last term in Eq. \eqref{eqomm} contains the two-electron Green's function, which in terms of $\CMcal{G}$ reads 
\begin{align} \label{eqG2}
 &  \frac{  \operatorname{Tr} \big( e^{-\beta \hat{K} }  \hat{\CMcal{T}}   \hat{\CMcal{S}}   \hat c_{ \sigma_\mathit{5}}^{ \dag}(\mathit{5}^{\smp})  \hat c_{ \sigma_\mathit{4}}^{}(\mathit{4}) \hat c_{\sigma_\mathit{3} }^{}(\mathit{3}) \hat c_{\sigma_\mathit{2}}^{ \dag}(\mathit{2})  \big)  }{ \operatorname{Tr} \big( e^{-\beta \hat{K} }   \hat{\CMcal{S}}   \big)} = \\
      & \hspace{0.2cm} -   \CMcal{G}_{\sigma_\mathit{3} \sigma_\mathit{2}}(\mathit{32})   \CMcal{G}_{\sigma_\mathit{4} \sigma_\mathit{5}}(\mathit{45}^{\smp}) 
      + 
       \frac{\delta \CMcal{G}_{\sigma_\mathit{3} \sigma_\mathit{2}}(\mathit{32})}{\delta \varphi_{\sigma_\mathit{5} \sigma_\mathit{4}}^{ext}(\mathit{5}^{\smp}\mathit{4}) }  , \nonumber
\end{align}   
as verified from the chain rule by differentiating Eq. \eqref{eqG}, where the $\varphi^{ext}$-dependence is contained in $\hat{\CMcal{S}}$ through Eq. \eqref{eqSa}. Inserting Eq. \eqref{eqG2} into Eq. \eqref{eqomm} shows that the term containing $\delta \CMcal{G}/\delta \varphi^{ext}$ complicates the access to a practically useful functional of $\CMcal{G}^{-1}$ in terms of $\CMcal{G}$. In Hedin's formalism \cite{firsthedin}, which does not impose a crossing symmetric interaction, the analogous term defines the self-energy $\Sigma$. But unlike Hedin's formalism, the $\delta \CMcal{G}/\delta \varphi^{ext}$-contribution to Eq. \eqref{eqomm} does not correspond to exchange and correlation, but to half the Hartree-Fock potential, plus correlations. The crossing symmetric starting point thus hints to a pathology of treating the two terms of Eq. \eqref{eqG2} asymmetrically, so we avoid introducing $\Sigma$. However, to invert Eq. \eqref{eqomm} it is unavoidable to use the chain rule $\delta \CMcal{G} = - \CMcal{G} \delta \CMcal{G}^{-1} \CMcal{G}$, so that both terms of Eq. \eqref{eqG2} contain a suitable factor of $\CMcal{G}$, yielding
\begin{align}
\CMcal{G}^{-1}_{\sigma_\mathit{1} \sigma_\mathit{2}}  \lp \mathit{12} \rp & = -   \big(  \delta_{\mathit{12}}^{}\partial_{t_\mathit{1}}^{}   +  k  (  \mathit{12} \rp \big) \delta_{\sigma_\mathit{1} \sigma_\mathit{2}}^{}     -    \varphi_{\sigma_\mathit{1} \sigma_\mathit{2}}^{tot} \lp \mathit{1^{\smp}2} \rp .
\end{align}
The total field splits into two components 
\begin{align}
\label{fields}
\varphi_{\sigma_\mathit{1} \sigma_\mathit{2}}^{tot} \lp \mathit{1^{\smp} 2} \rp & = \varphi_{\sigma_\mathit{1} \sigma_\mathit{2}}^{ext} \lp \mathit{1^{\smp} 2} \rp  + \varphi_{\sigma_\mathit{1} \sigma_\mathit{2}}^{ind}\lp \mathit{1^{\smp} 2} \rp ,
\end{align}
where the induced field reads 
\begin{align} \label{inducedfield}
\varphi_{\sigma_\mathit{1} \sigma_\mathit{2}}^{ind} \lp \mathit{1^{\smp} 2} \rp  
  =  v^{\sigma_\mathit{1} \sigma_\mathit{2}}_{\sigma_\mathit{3} \sigma_\mathit{4}} &
\lp \mathit{12},\! \mathit{34} \rp  \CMcal{G}_{\sigma_\mathit{3} \sigma_\mathit{4}} \lp \mathit{34^{\smp}} ) \\
   -    v^{\sigma_\mathit{1} \sigma_\mathit{3}}_{\sigma_\mathit{4} \sigma_\mathit{5}}
(\mathit{13} ^{\!},^{\!} \mathit{45})    \CMcal{G}_{\sigma_\mathit{3} \sigma_\mathit{6}} & (\mathit{36^{\smp}}) \frac{\delta \varphi_{\sigma_\mathit{6} \sigma_\mathit{2}}^{tot}(\mathit{6^{\smp}2 } )  }{\delta \varphi_{\sigma_\mathit{5} \sigma_\mathit{4}}^{ext}(\mathit{ 5^{\smp}4 }) } \nonumber  .
\end{align}
This is the ``mass operator'' in Hedin's theory. Since this is varied when deriving the spin susceptibility, the Green's function has to be kept spin-offdiagonal until the end, despite the absence of spin-orbit coupling. Before continuing along these lines, we turn to the four-vector representation in \ref{secD}, where the connection to the macroscopic Maxwell fields is clarified. 
\subsection{Four-vector representation} \label{secD}
The external, induced and total fields can all be written on the generic Pauli form 
\begin{align} \label{varphi1}
\varphi_{\sigma_\mathit{1} \sigma_\mathit{2}}^{}  ( \mathit{1^{\smp} 2} )    & = \sigma^\mu_{ \sigma_\mathit{1} \sigma_\mathit{2} } \varphi_\mu^{} ( \mathit{1^{\smp} 2} )     , \\ \label{varphi2}
\varphi_\mu^{} ( \mathit{1^{\smp} 2} )    & =  \tfrac{1}{2}\sigma^\mu_{ \sigma_\mathit{4} \sigma_\mathit{3} }   \varphi_{\sigma_\mathit{3} \sigma_\mathit{4}}^{}( \mathit{1^{\smp} 2} ) ,
\end{align}
with four-vector index $\mu=0,x,y,z$, and  
\begin{align}
\varphi_\mu^{} ( \mathit{1^{\smp} 2} )  & = \big( \phi ( \mathit{1^{\smp} 2} )  , - \tfrac{1}{2} B_i^{} ( \mathit{1^{\smp} 2} )  \big)     .
\end{align}
Here, $\phi$ is the electric scalar potential ($E_i = - \partial_i \phi$), $B_i$ the magnetic flux density and $\tfrac{1}{2}$ the Bohr magneton. To be consistent with Eqs. \eqref{varphi1} and \eqref{varphi2}, field derivatives must be related as 
\begin{align}  \label{fd1} 
\frac{\delta }{\delta \varphi_{\sigma_\mathit{1} \sigma_\mathit{2}}^{}  ( \mathit{1^{\smp} 2} )   }  & = \tfrac{1}{2} \sigma^\mu_{ \sigma_\mathit{2} \sigma_\mathit{1} } \frac{\delta  }{ \varphi_\mu^{} ( \mathit{1^{\smp} 2} )   }   , \\ 
\label{fd2}
\frac{\delta }{\delta \varphi_\mu^{} ( \mathit{1^{\smp} 2} ) }   & =  \sigma^\mu_{ \sigma_\mathit{3} \sigma_\mathit{4} }  \frac{\delta }{\delta  \varphi_{\sigma_\mathit{3} \sigma_\mathit{4}}^{}( \mathit{1^{\smp} 2} ) }  .
\end{align}  
The spin-density matrix $\varrho$, which is related to the Green's function through the simple relation
\begin{align}
\label{n}
\varrho_{\sigma_\mathit{1} \sigma_\mathit{2}} ( \mathit{1^{\smp} 2} ) & = \CMcal{G}_{\sigma_\mathit{2} \sigma_\mathit{1}} ( \mathit{2 1^{\smp} } ),
\end{align} 
has the same Pauli form as field derivatives, namely 
\begin{align} \label{G1}
\varrho_{\sigma_\mathit{1} \sigma_\mathit{2}} ( \mathit{1^{\smp} 2} )  & =  \tfrac{1}{2} \sigma^\mu_{\sigma_\mathit{2} \sigma_\mathit{1}} \varrho_\mu^{}  ( \mathit{1^{\smp} 2} )    , \\
\varrho_\mu^{} ( \mathit{1^{\smp} 2} )   & =  \sigma^\mu_{\sigma_\mathit{3} \sigma_\mathit{4}} \varrho_{\sigma_\mathit{3} \sigma_\mathit{4}}^{} ( \mathit{1^{\smp} 2} )  ,
\end{align}
where, notably, $\varrho_0=n$ is the electronic density and $\varrho_z=m$ is the spin magnetization. Likewise, the spin-density derivatives are on the same form as the fields,
\begin{align}
\frac{\delta }{\delta  \varrho_{\sigma_\mathit{1} \sigma_\mathit{2}} ( \mathit{1^{\smp} 2} ) }  & = \sigma^\mu_{ \sigma_\mathit{1} \sigma_\mathit{2} }\frac{\delta }{ \delta  \varrho_{\mu} ( \mathit{1^{\smp} 2} ) } , \\
\frac{\delta }{ \delta  \varrho_{\mu} ( \mathit{1^{\smp} 2} ) } & =  \tfrac{1}{2}\sigma^\mu_{ \sigma_\mathit{4} \sigma_\mathit{3} }  \frac{\delta }{\delta \varrho_{ \sigma_\mathit{3} \sigma_\mathit{4} } ( \mathit{1^{\smp} 2} )  }  .
\end{align}
The four-vector representation of Eq. \eqref{inducedfield} is obtained from Eqs. \eqref{eqvnajs}, \eqref{eqv0}, \eqref{varphi2}, \eqref{fd2}, \eqref{n} and \eqref{G1}, yielding
\begin{align} \label{indnew}
\varphi_{\mu_\mathit{1}}^{ind} \lp \mathit{1^{\smp} 2} \rp  
  =    v_{\mu_\mathit{1} \mu_\mathit{3}} & (\mathit{12},\mathit{34})  \varrho_{\mu_\mathit{3}}   \lp \mathit{4^{\smp} 3} )   \\
  -  \tfrac{1}{4  } ^{\!} \operatorname{tr} {^{\!}} \big( \sigma^{\mu_\mathit{1}}  {^{\!}} \sigma^{\mu_\mathit{2}}   {^{\!}} \sigma^{\mu_\mathit{3}}  {^{\!}} \sigma^{\mu_\mathit{4}}   \big) v_{\mu_\mathit{2} \mu_\mathit{5}} & (\mathit{13},\mathit{45}) 
  \varrho_{\mu_\mathit{3}} \lp \mathit{6^{\smp} 3} ) 
  \frac{\delta \varphi^{tot}_{\mu_\mathit{4}} (\mathit{6^{\smp}2 } )  }{\delta \varphi_{\mu_\mathit{5}}^{ext}(\mathit{ 5^{\smp}4 }) }    \nonumber  ,
\end{align} 
where $ \operatorname{tr} \big( \sigma^{\mu_\mathit{1}} \sigma^{\mu_\mathit{2}} \big)  = 2 \delta_{\mu_\mathit{1} \mu_\mathit{2}} $ has been used twice in the first term. The trace in the second term can be evaluated, but contains many terms. This shows the benefit of the usual ``spin representation'' --- another is that it naturally expresses the crossing symmetry, like in Eq. \eqref{crossing}. Since this will be relevant later when deriving the spin-flip interaction, we will return to the spin representation, but we first stress the correspondence to field quantities in the macroscopic Maxwell theory, reading \cite{starke} 
\begin{align}
\varphi^{tot}_\mu     & = \big( \phi , - \tfrac{1}{2} B_i^{}      \big)   , \\
\varphi^{ext}_\mu      & =  \big( \tfrac{1}{\varepsilon_0} \phi_{D}^{}   , - \tfrac{1}{2} \mu_0 H_i^{}     \big)   , \\
\varphi^{ind}_\mu   & = \big( \!-\! \tfrac{1}{\varepsilon_0}  \phi_P^{} , - \tfrac{1}{2}  \mu_0  M_i^{} \big)   ,  
\end{align}
where $\phi_D^{}$ and $\phi_P^{}$ are scalar potentials for the electric displacement and polarization fields, assumed to be conservative, i.e. $D_i^{} = - \partial_i \phi_D^{}$ and $P_i = - \partial_i \phi_P^{}$. In addition, $H_i^{}$ is the inductive magnetic field and $M_i^{}$ the magnetization field. For clarity, we have refrained from utilizing that $\varepsilon_0 = 1/4\pi$ and $\mu_0 = 4\pi \alpha^2 $ in atomic units, where $\alpha \approx 1/137$ is the fine-structure constant. With this correspondence, it is easy to identify the inverse relative dielectric function as well as relative permeability, as
\begin{align}
 \varepsilon^{\mathsmaller{-1}}_r(\mathit{12},\mathit{34}) & =  \frac{\delta \varphi^{tot}_0 ( \mathit{1^{\smp} 2} ) }{\delta \varphi^{ext}_0(\mathit{3}^{\smp} \mathit{4})} ,   \\
 \label{eqmu}
{\mu_r}_{ij} (\mathit{12},\mathit{34}) & =  \frac{\delta \varphi^{tot}_i ( \mathit{1^{\smp} 2} ) }{\delta  \varphi^{ext}_j(\mathit{3}^{\smp} \mathit{4})}   .
\end{align}
There are also components $\delta \varphi^{tot}_0/\delta \varphi_j^{ext}$ and $\delta \varphi_i^{tot}/\delta \varphi_0^{ext}$, which are important in bianisotropic media and multiferroics. Common to all components is that they describe an ``inverse electromagnetic screening factor'' $\mathcal{S}$. We therefore define
\begin{align}
 \mathcal{S}^{\mathsmaller{-1}}_{\mu_\mathit{1} \mu_\mathit{3}}(\mathit{12},\mathit{34}) & =  \frac{\delta \varphi^{tot}_{\mu_\mathit{1}} ( \mathit{1^{\smp} 2} ) }{\delta \varphi^{ext}_{\mu_\mathit{3}}(\mathit{3}^{\smp} \mathit{4})}  .
\end{align} 
The static limit of the charge-charge and spin-spin components are small in metals and diamagnets, respectively, whereas in ferromagnets the latter instead diverges.
\subsection{Screened collective four-point interaction}  \label{sfi}
Having established connections to Maxwell's theory, we continue on the path towards the magnon-phonon coupling. In Eq. \eqref{inducedfield}, the main quantity to find is  
\begin{align}
 {\mathcal{S}^{\mathsmaller{-1}}}|^{\sigma_\mathit{1} \sigma_\mathit{2}}_{\sigma_\mathit{3} \sigma_\mathit{4}}(\mathit{12},\mathit{34})  & = \frac{\delta \varphi^{tot}_{\sigma_\mathit{1} \sigma_\mathit{2}}( \mathit{1^{\smp} 2} )}{\delta \varphi^{ext}_{\sigma_\mathit{3} \sigma_\mathit{4}} ( \mathit{3^{\smp} 4} )}  .
\end{align}
Using Eq. \eqref{fields} and the chain rule yields  
\begin{align}
   \mathcal{S}^{\mathsmaller{-1} }|^{\sigma_\mathit{1} \sigma_\mathit{2}}_{\sigma_\mathit{3} \sigma_\mathit{4}}(\mathit{12},\mathit{34})  &   =   \delta_{\sigma_\mathit{1} \sigma_\mathit{3}}^{} \delta_{\sigma_\mathit{2} \sigma_\mathit{4}}^{} \delta_{\mathit{13}}^{} \delta_{\mathit{24}}^{}  \\
 + 
\mathcal{V}^{\sigma_\mathit{1} \sigma_\mathit{2}}_{\sigma_\mathit{5} \sigma_\mathit{6}}(\mathit{12},\mathit{56}) 
 \mathcal{P}^{\sigma_\mathit{5} \sigma_\mathit{6}}_{\sigma_\mathit{7} \sigma_\mathit{8}} & (\mathit{56}    ,  \mathit{78})
{\mathcal{S}^{\mathsmaller{-1}}}|^{\sigma_\mathit{7} \sigma_\mathit{8}}_{\sigma_\mathit{3} \sigma_\mathit{4}}(\mathit{78},\mathit{34})     , 
  \nonumber 
\end{align}
where 
\begin{align} \label{int}
\mathcal{V}^{\sigma_\mathit{1} \sigma_\mathit{2}}_{\sigma_\mathit{3} \sigma_\mathit{4}}(\mathit{12},\mathit{34})  =  \frac{\delta \varphi^{ind}_{\sigma_\mathit{1} \sigma_\mathit{2}}( \mathit{  \mathit{1^{\smp} 2}  })}{\delta \varrho_{\sigma_\mathit{4} \sigma_\mathit{3}}^{}(\mathit{4^{\smp} 3 } ) } 
\end{align}
is the {\it collective} four-point interaction, which in Hedin's formalism amounts to the sum of Coulomb interaction and irreducible four-point vertex $\delta \Sigma/\delta \CMcal{G}$ \cite{reining2002,kutepov2017}, and 
\begin{align}   \label{P1}
\mathcal{P}^{\sigma_\mathit{1} \sigma_\mathit{2}}_{\sigma_\mathit{3} \sigma_\mathit{4}}(\mathit{12},\mathit{34}) & = \frac{\delta \varrho_{\sigma_\mathit{2} \sigma_\mathit{1}}^{}(\mathit{ 2^{\smp} 1 } ) }{\delta \varphi^{tot}_{\sigma_\mathit{3} \sigma_\mathit{4}}( \mathit{ 3^{\smp} 4  } ) }  \\ \label{P2}
& = \CMcal{G}_{\sigma_\mathit{1} \sigma_\mathit{3}}^{}(\mathit{ 13^{\smp} } ) \CMcal{G}_{\sigma_\mathit{4} \sigma_\mathit{2}}^{}(\mathit{ 42^{\smp} } ) 
\end{align}
is the {\it free} electron-hole pair propagator, which might also be called the four-point electromagnetic polarization function. By defining the {\it screened collective} four-point interaction 
\begin{align} \label{eqW}
\mathcal{W}^{\sigma_\mathit{1} \sigma_\mathit{2}}_{\sigma_\mathit{3} \sigma_\mathit{4}}(\mathit{12},\mathit{34})  &  =  {\mathcal{S}^{\mathsmaller{-1}}}|^{\sigma_\mathit{1} \sigma_\mathit{2}}_{\sigma_\mathit{5} \sigma_\mathit{6}}(\mathit{12},\mathit{56}) \mathcal{V}^{\sigma_\mathit{5} \sigma_\mathit{6}}_{\sigma_\mathit{3} \sigma_\mathit{4}}(\mathit{56},\mathit{34}) ,
\end{align}
it is possible to write 
\begin{align} \label{Sinv}
   \mathcal{S}^{\mathsmaller{-1} }|^{\sigma_\mathit{1} \sigma_\mathit{2}}_{\sigma_\mathit{3} \sigma_\mathit{4}}(\mathit{12},\mathit{34})    &  =    \delta_{\sigma_\mathit{1} \sigma_\mathit{3}}^{} \delta_{\sigma_\mathit{2} \sigma_\mathit{4}}^{} \delta_{\mathit{13}}^{} \delta_{\mathit{24}}^{}    \\
 + 
\mathcal{W}^{\sigma_\mathit{1} \sigma_\mathit{2}}_{\sigma_\mathit{5} \sigma_\mathit{6}}(\mathit{12}, \mathit{5} & \mathit{6}  ) 
  \mathcal{P}^{\sigma_\mathit{5} \sigma_\mathit{6}}_{\sigma_\mathit{3} \sigma_\mathit{4}}  (\mathit{56}    ,  \mathit{34}) ,
  \nonumber 
\end{align}
which when plugged into Eq. \eqref{inducedfield} yields 
\begin{align}   \label{ind}
\varphi_{\sigma_\mathit{1} \sigma_\mathit{2}}^{ind} \lp \mathit{1^{\smp} 2} \rp  
  &  =  2  v^{\sigma_\mathit{1} \sigma_\mathit{2}}_{\sigma_\mathit{3} \sigma_\mathit{4}} 
\lp \mathit{12},\! \mathit{34} \rp  \CMcal{G}_{\sigma_\mathit{3} \sigma_\mathit{4}}^{} \lp \mathit{34^{\smp}} )  
 \\
  -    v^{\sigma_\mathit{1} \sigma_\mathit{3}}_{\sigma_\mathit{4} \sigma_\mathit{5}}
(\mathit{13} ^{\!},^{\!} \mathit{45})     & \CMcal{G}_{\sigma_\mathit{3} \sigma_\mathit{6}}^{} (\mathit{36^{\smp}})   \CMcal{G}_{\sigma_\mathit{7} \sigma_\mathit{5}}^{}(\mathit{75^{\smp}}) \CMcal{G}_{\sigma_\mathit{4} \sigma_\mathit{8}}^{}(\mathit{48^{\smp}})   \nonumber
   \\
  & \smalltimes \!   \mathcal{W}^{\sigma_\mathit{6} \sigma_\mathit{2}}_{\sigma_\mathit{7} \sigma_\mathit{8}}(\mathit{62}, \mathit{7}  \mathit{8}  ) 
\nonumber  .
\end{align}
The first term is the Hartree-Fock potential, where the factor of $2$ is a consequence of the crossing symmetry of $v$ in Eq. \eqref{crossing}. The second term has employed Eq. \eqref{P2} and contains all correlations through $\mathcal{W}$, which like $v$, fulfils the crossing symmetry 
\begin{align} \label{crossW}
\mathcal{W}^{\sigma_\mathit{1} \sigma_\mathit{2}}_{\sigma_\mathit{3} \sigma_\mathit{4}}(\mathit{12}, \mathit{3}  \mathit{4}  )  & =  - \mathcal{W}^{\sigma_\mathit{1} \sigma_\mathit{3}}_{\sigma_\mathit{2} \sigma_\mathit{4}}(\mathit{13}, \mathit{2}  \mathit{4}  ) ,
\end{align}
or likewise if $\mathit{1} \sigma_\mathit{1}$ and $\mathit{4} \sigma_{\mathit{4}}$ are interchanged. This guarantees that the wavefunction is antisymmetric when interchanging electrons, and is proven by repeating the derivation when initially anticommuting the annihilation operators in Eq. \eqref{eqKint}. $\mathcal{W}$ has another symmetry related to pair-hopping, proven by using anomalous probing fields which interchange electrons and positrons, but this is of little interest in this work. 
\subsection{Spin susceptibility and spin-flip interaction}
The {\it interacting} electron-hole propagator is defined as  
\begin{align}   \label{firstR}
\mathcal{R}^{\sigma_\mathit{1} \sigma_\mathit{2}}_{\sigma_\mathit{3} \sigma_\mathit{4}}(\mathit{12},\mathit{34}) & = \frac{\delta \varrho_{\sigma_\mathit{2} \sigma_\mathit{1}}^{}(\mathit{ 2^{\smp} 1 } ) }{\delta \varphi^{ext}_{\sigma_\mathit{3} \sigma_\mathit{4}}( \mathit{ 3^{\smp} 4  } ) }  ,
\end{align}
which is analogous to $\mathcal{P}$ in Eq. \eqref{P1} but with $\varphi^{tot}$ replaced by $\varphi^{ext}$. The physical charge and spin susceptibilities are obtained by contraction, through  
\begin{align} \label{susc}
 \mathcal{R}^{\sigma_\mathit{1} \sigma_\mathit{2}}_{\sigma_\mathit{3} \sigma_\mathit{4}}(\mathit{13})  & =  \mathcal{R}^{\sigma_\mathit{1} \sigma_\mathit{2}}_{\sigma_\mathit{3} \sigma_\mathit{4}}(\mathit{11},\mathit{33}),
\end{align}
which, with Eqs. \eqref{fd1}, \eqref{n} and \eqref{G1} can be expressed as 
\begin{align}  \label{sjsj}
\mathcal{R}^{\sigma_\mathit{1} \sigma_\mathit{2}}_{\sigma_\mathit{3} \sigma_\mathit{4}}(\mathit{13}) & =  \frac{   \sigma^{\mu}_{\sigma_\mathit{1} \sigma_\mathit{2}} }{2}    \frac{\delta \varrho_{ \mu }^{}(\mathit{ 1^{\smp} 1 } ) }{\delta \varphi^{ext}_{ \nu }( \mathit{ 3^{\smp} 3  } ) }  \frac{\sigma^{\nu}_{\sigma_\mathit{4} \sigma_\mathit{3}} }{2}   .
\end{align} 
Since $z$ is the spin direction, Eq. \eqref{sjsj} shows that the transverse spin fluctuations are contained in $\mathcal{R}^{\downarrow \! \uparrow}_{\downarrow \! \uparrow}$ and $\mathcal{R}^{\uparrow \! \downarrow}_{\uparrow \! \downarrow}$, which yield the magnon spectra through their imaginary parts. Since the two are related through inversion, we will consider only the former. From \eqref{firstR} and the chain rule, we get 
\begin{align} \label{eqgreat}
  &  \mathcal{R}^{\downarrow \! \uparrow}_{\downarrow \! \uparrow}(\mathit{12},\mathit{34})  = \mathcal{P}^{\downarrow \! \uparrow}_{\downarrow \! \uparrow}(\mathit{12},\mathit{34}) \\
  + \mathcal{P}^{\downarrow \! \uparrow}_{\downarrow \! \uparrow} & ( \mathit{12}, \mathit{56}) \mathcal{V}^{\downarrow \! \uparrow}_{\downarrow \! \uparrow}(\mathit{56},\mathit{78})  \mathcal{R}^{\downarrow \! \uparrow}_{\downarrow \! \uparrow}(\mathit{78},\mathit{34})   . \nonumber  
\end{align}
With spin-orbit coupling, this equation would, due to a spin-nondiagonal $\CMcal{G}$, also couple charge- and spin-components of $\mathcal{R}$, and when accounting for phonons, contain the process of magnon-phonon interconversion \cite{ruckriegel2014,Streib2019}. This could be solved at the level of the random-phase approximation. In this work, we focus on the non-relativistic magnon number-conserving processes. The goal is to derive contributions to the {\it spin-flip} interaction $\mathcal{V}^{\downarrow \! \uparrow}_{\downarrow \! \uparrow}$  describing the simultaneous propagation of spin and charge excitations, which allow for spin conservation. Since this is a four-point interaction, it is necessary to solve for the full four-point $\mathcal{R}^{\downarrow \! \uparrow}_{\downarrow \! \uparrow}$ and contract it afterwards according to Eq. \eqref{susc}, to access the spin-susceptibility. 
\section{Approximate theory}
\subsection{Low-energy model}
The long-range interaction $v$, which enters the spin-flip interaction through Eqs. \eqref{int} and \eqref{ind}, makes calculations very expensive. Trading rigour for clarity, we assume that a renormalized low-energy model has properly been constructed in advance, and make the replacement  
\begin{align} \label{eqv199}
    v^{\sigma_\mathit{1} \sigma_\mathit{2}}_{\sigma_\mathit{3} \sigma_\mathit{4}}(\mathit{12},\mathit{34}) &  \coloneqq \mathcal{U}^{\sigma_\mathit{1} \sigma_\mathit{2}}_{\sigma_\mathit{3} \sigma_\mathit{4}}(\mathit{1}) \delta_{\mathit{12}}^{} \delta_{\mathit{23}}^{} \delta_{\mathit{34}}^{}  \\ \label{eqU20}
 \mathcal{U}^{\sigma_\mathit{1} \sigma_\mathit{2}}_{\sigma_\mathit{3} \sigma_\mathit{4}}(\mathit{1})   & =  \frac{   \mathcal{U}_{n_\mathit{1}}^{} }{2}  \big( \delta_{\sigma_\mathit{1} \sigma_\mathit{2}}^{} \delta_{\sigma_\mathit{3} \sigma_\mathit{4}}^{}   -    \delta_{\sigma_\mathit{1} \sigma_\mathit{3}}^{} \delta_{\sigma_\mathit{2} \sigma_\mathit{4}}^{} \big),
\end{align}
which assumes an instantaneous, local and orbital-diagonal interaction, which still fulfils crossing symmetry. This neglects retardation, couplings between different unit cells and between different orbitals in the same unit cell, and is justified if the low-energy electronic structure has a single isolated spin-resolved band at the Fermi energy. This step is easily avoided if needed, but is instructive for investigating the magnon-phonon coupling, owing to the fact that a single magnon branch only requires a one-band model. 

Since the spin-flip interaction is obtained by varying $\varphi^{ind}_{\downarrow \! \uparrow}$, we insert the replacement of Eq. \eqref{eqv199} into Eq. \eqref{ind}, and get $\mathcal{w}$
\begin{align}   \label{lax}
 \varphi_{\downarrow \! \uparrow}^{ind} \lp \mathit{1^{\smp} 2} \rp  
    =  2 \delta_{\mathit{12}}^{}  &   \mathcal{U}^{\downarrow \! \uparrow}_{\downarrow \! \uparrow} 
\lp \mathit{1} \rp  \CMcal{G}_{\downarrow \! \uparrow}^{} \lp \mathit{22^{\smp}} )     
 \\
  -    \mathcal{U}^{\downarrow  \sigma_\mathit{3}}_{\sigma_\mathit{4} \sigma_\mathit{5}}
\lp \mathit{1} \rp \CMcal{G}_{\sigma_\mathit{3} \sigma_\mathit{6}}^{} (\mathit{16^{\smp}})   \CMcal{G}_{\sigma_\mathit{7} \sigma_\mathit{5}}^{} (& \mathit{71^{\smp}}) \CMcal{G}_{\sigma_\mathit{4} \sigma_\mathit{8}}^{}(\mathit{18^{\smp}})       \mathcal{W}^{\sigma_\mathit{6} \uparrow}_{\sigma_\mathit{7} \sigma_\mathit{8}}(\mathit{62}, \mathit{7}  \mathit{8}  ) 
\nonumber  ,
\end{align} 
which, when used in Eq. \eqref{int}, yields 
\begin{align} \label{inter}
\mathcal{V}^{\downarrow \! \uparrow}_{\downarrow \! \uparrow}(\mathit{12},\mathit{34})  = 2 & \delta_{\mathit{12}}^{}   \delta_{\mathit{23}}^{} \delta_{\mathit{34}}^{} \mathcal{U}^{\downarrow \! \uparrow}_{\downarrow \! \uparrow} 
\lp \mathit{1} \rp   
 \\
  - 2 \delta_{\mathit{13}}   \mathcal{U}^{\downarrow  \! \downarrow}_{\uparrow \! \uparrow}
\lp \mathit{1} \rp  \CMcal{G}_{\uparrow }^{}(\mathit{15^{\smp}}) & \CMcal{G}_{  \uparrow}^{} ( \mathit{61^{\smp}})       \mathcal{W}^{\uparrow \! \uparrow}_{\uparrow \! \uparrow}(\mathit{42}, \mathit{6 5}  )  \nonumber   \\[-3pt]
 +  2\delta_{\mathit{14}}  \mathcal{U}^{\downarrow \! \uparrow}_{\downarrow  \! \uparrow}
\lp \mathit{1} \rp \CMcal{G}_{ \downarrow  }^{} (\mathit{15^{\smp}}) &   \CMcal{G}_{ \uparrow  }^{}(\mathit{16^{\smp}})       \mathcal{W}^{ \downarrow \! \uparrow}_{\downarrow \!  \uparrow}(\mathit{52}, \mathit{36}  ) \nonumber   \\[-3pt]
 -  2  \mathcal{U}^{\downarrow \! \downarrow}_{\uparrow \! \uparrow}
\lp \mathit{1} \rp \CMcal{G}_{ \downarrow  }^{} (\mathit{16^{\smp}})   \CMcal{G}_{  \uparrow }^{} ( \mathit{7 1^{\smp}} & )     \CMcal{G}_{ \uparrow }^{}(\mathit{18^{\smp}})       \frac{\delta  \mathcal{W}^{\downarrow  \! \uparrow}_{ \uparrow \!  \uparrow }(\mathit{62}, \mathit{7}  \mathit{8}  ) }{\delta \CMcal{G}_{\downarrow \! \uparrow}^{}(\mathit{34}^{\smp})}    
 \nonumber ,
\end{align}
where the crossing symmetry of $\mathcal{U}$ and $\mathcal{W}$ have been used as well as the spin-diagonality of $\CMcal{G}$, valid in the absence of spin-orbit coupling. $\mathcal{U}$ has been treated as $\CMcal{G}$-independent, in analogy with $v$ in the full Hilbert space. The functional derivative $\delta \mathcal{W}/\delta \CMcal{G}$ will be shown to contain the coupling between spin and charge fluctuations, and is the term that, when generalized to dynamical lattices, will contain the magnon-phonon coupling. The generalization of Eq. \eqref{inter} for arbitrary spin-components of $\mathcal{V}$ together with Eq. \eqref{eqW} for $\mathcal{W}$ form a self-consistent set, even when $\CMcal{G}$ and $\mathcal{P}$ are treated as predetermined inputs, for example from density functional theory. Since this is beyond the reach of present-day computational capacities, we will solve Eq. \eqref{inter} iteratively. 
\subsection{Iterative spin-flip interaction} \label{secit}
A practical scheme is iterative rather than self-consistent. The first iteration is obtained by dropping $\mathcal{W}$ in the right-hand side of Eq. \eqref{inter}. Using Eq. \eqref{eqU20}, the first iteration yields
\begin{align} \label{inter1}
{\mathcal{V}^{\bcdot }}^{\downarrow \! \uparrow}_{\downarrow \! \uparrow}\lp \mathit{12},\mathit{34} \rp & =       {\delta^{\!}}_{\mathit{12}}^{}    {\delta^{\!}}_{\mathit{34}}^{}   {\mathcal{V}^{\bcdot }}^{\downarrow \! \uparrow}_{\downarrow \! \uparrow}\lp \mathit{13} \rp =    -  {\delta^{\!}}_{\mathit{12}}^{}   {\delta^{\!}}_{\mathit{23}}^{} {\delta^{\!}}_{\mathit{34}}^{}  \mathcal{U}_{n_\mathit{1}}^{} ,
\end{align}
where the number of $\bcdot$'s denotes the iteration number. The one-point structure, which also holds for the other spin-components of $\mathcal{V}^{\bcdot}$, implies that $\mathcal{W}^{\bcdot}$ has the same two-point structure for all spin-components. Since this clearly breaks the crossing symmetry of $\mathcal{W}^{\bcdot}$, we use a trick where $\mathcal{W}_c^{\bcdot}= \mathcal{W}^{\bcdot} - \mathcal{V}^{\bcdot} = \mathcal{W}^{\bcdot} \mathcal{P} \mathcal{V}^{\bcdot}$ is approximated as
\begin{align} \label{naajs}
    {\mathcal{W}_c^{\bcdot}}^{ \sigma_\mathit{1} \sigma_\mathit{2}}_{\sigma_\mathit{3} \sigma_\mathit{4}}(\mathit{12},\mathit{34}) & \approx
    \delta_{\mathit{12}}^{}\delta_{\mathit{34}}^{} {\mathcal{W}_c^{\bcdot}}^{ \sigma_\mathit{1} \sigma_\mathit{2}}_{\sigma_\mathit{3} \sigma_\mathit{4}}(\mathit{13})
    \\    & - \delta_{\mathit{13}}^{} \delta_{\mathit{24}}^{} 
   {\mathcal{W}_c^{\bcdot}}^{ \sigma_\mathit{1} \sigma_\mathit{3}}_{\sigma_\mathit{2} \sigma_\mathit{4}}(\mathit{12}) , \nonumber 
\end{align}
where we, for convenience, denote the two-point interactions as $\mathcal{W}_c^{\bcdot}$, although their physical dimensions differ from the four-point interactions by two factors of time. Equation \eqref{naajs} corresponds to keeping direct and exchange matrix elements, but we have used the crossing symmetry on the latter. Notice that the double counting of the component $ {\mathcal{W}_c^{\bcdot}} {\color{red} |}^{\sigma_\mathit{1} \sigma_\mathit{2}}_{\sigma_\mathit{3} \sigma_\mathit{4}}(\mathit{11})$ introduces no error, unless the imaginary time is discretized. Since our goal is to express the next iteration interaction, $\mathcal{V}^{\bcdot \bcdot}$, in terms of $\CMcal{G}$, $\mathcal{V}^{\bcdot}$ and $\mathcal{W}_c^{\bcdot}$, it is clear from Eqs. \eqref{inter}, \eqref{inter1} and \eqref{naajs} that it only remains to find expressions for the factors $\delta \mathcal{W}^{\bcdot}/\delta \CMcal{G}$, with $\mathcal{W}^{\bcdot}$ corresponding to the two-point reductions in Eq. \eqref{naajs}. For arbitrary spin components, it follows from Eqs. \eqref{eqW} and \eqref{Sinv} that $\mathcal{W}^{{\bcdot} \mathsmaller{-1} } = \mathcal{V}^{ {\bcdot} \mathsmaller{-1} } - \mathcal{P}$, and consequently from the chain rule
\begin{align} \label{dW}
\frac{\delta {\mathcal{W}^{\bcdot ^{\!}}  }^{\sigma_\mathit{1} \sigma_\mathit{2}}_{\sigma_\mathit{3} \sigma_\mathit{4}} \lp \mathit{13} \rp }{\delta \CMcal{G}_{\sigma_\mathit{5} \sigma_\mathit{6}} \lp \mathit{56^{\smp}} \rp} & ^{\!} = ^{\!} {\mathcal{W}^{\bcdot ^{\!} }}^{\sigma_\mathit{1} \sigma_\mathit{2}}_{\sigma_\mathit{7} \sigma_\mathit{8}} \lp \mathit{17} \rp  \frac{\delta \mathcal{P}^{\sigma_\mathit{7} \sigma_\mathit{8}}_{\sigma_\mathit{9} \sigma_\mathit{10}}\lp \mathit{79} \rp }{\delta \CMcal{G}_{\sigma_\mathit{5} \sigma_\mathit{6}}\lp \mathit{56^{\smp}} \rp} 
{\mathcal{W}^{\bcdot ^{\!} }}^{\sigma_\mathit{9} \sigma_\mathit{10}}_{\sigma_\mathit{3} \sigma_\mathit{4}} \lp \mathit{93} \rp 
 ,
\end{align} 
where the $\CMcal{G}$-independence of $\mathcal{V}^{\bcdot}$ in Eq. \eqref{inter1} has been used. The two factors of $\mathcal{W}^{\bcdot}$ in the right-hand side do in general have three-point contributions, but these must vanish due to Eq. \eqref{naajs}. From Eq. \eqref{P2} it follows that 
\begin{align} \label{dP}
 \frac{\delta \mathcal{P}^{\sigma_\mathit{7} \sigma_\mathit{8}}_{\sigma_\mathit{9} \sigma_\mathit{10}}\lp \mathit{79} \rp }{\delta \CMcal{G}_{\sigma_\mathit{5} \sigma_\mathit{6}}^{}\lp \mathit{56^{\smp}} \rp}  & = \delta_{\sigma_\mathit{7} \sigma_\mathit{5}}^{} \delta_{\sigma_\mathit{9} \sigma_\mathit{6}}^{} \delta_{\mathit{75}}^{} \delta_{\mathit{96}}^{}
   \CMcal{G}_{\sigma_\mathit{10} \sigma_\mathit{8}}^{} \lp \mathit{ 97^{\smp} } )  \\[-5pt]
 & +  \delta_{\sigma_\mathit{10} \sigma_\mathit{5}}^{} \delta_{\sigma_\mathit{8} \sigma_\mathit{6}}^{} \delta_{\mathit{95}}^{} \delta_{\mathit{76}}^{} \CMcal{G}_{\sigma_\mathit{7} \sigma_\mathit{9}}^{} \lp \mathit{ 79^{\smp} } )  \nonumber .
\end{align}
Using Eqs. \eqref{P2}, \eqref{eqv199}, \eqref{eqU20}, \eqref{inter1}, \eqref{naajs}, \eqref{dW} and \eqref{dP}, the second iteration of the spin-flip interaction in Eq. \eqref{inter}, which is obtained by treating screening perturbatively through the approximation $\mathcal{W} \approx \mathcal{W}^{\bcdot}$, takes the form
\begin{figure}[h!]
\includegraphics[width=1\linewidth,right]{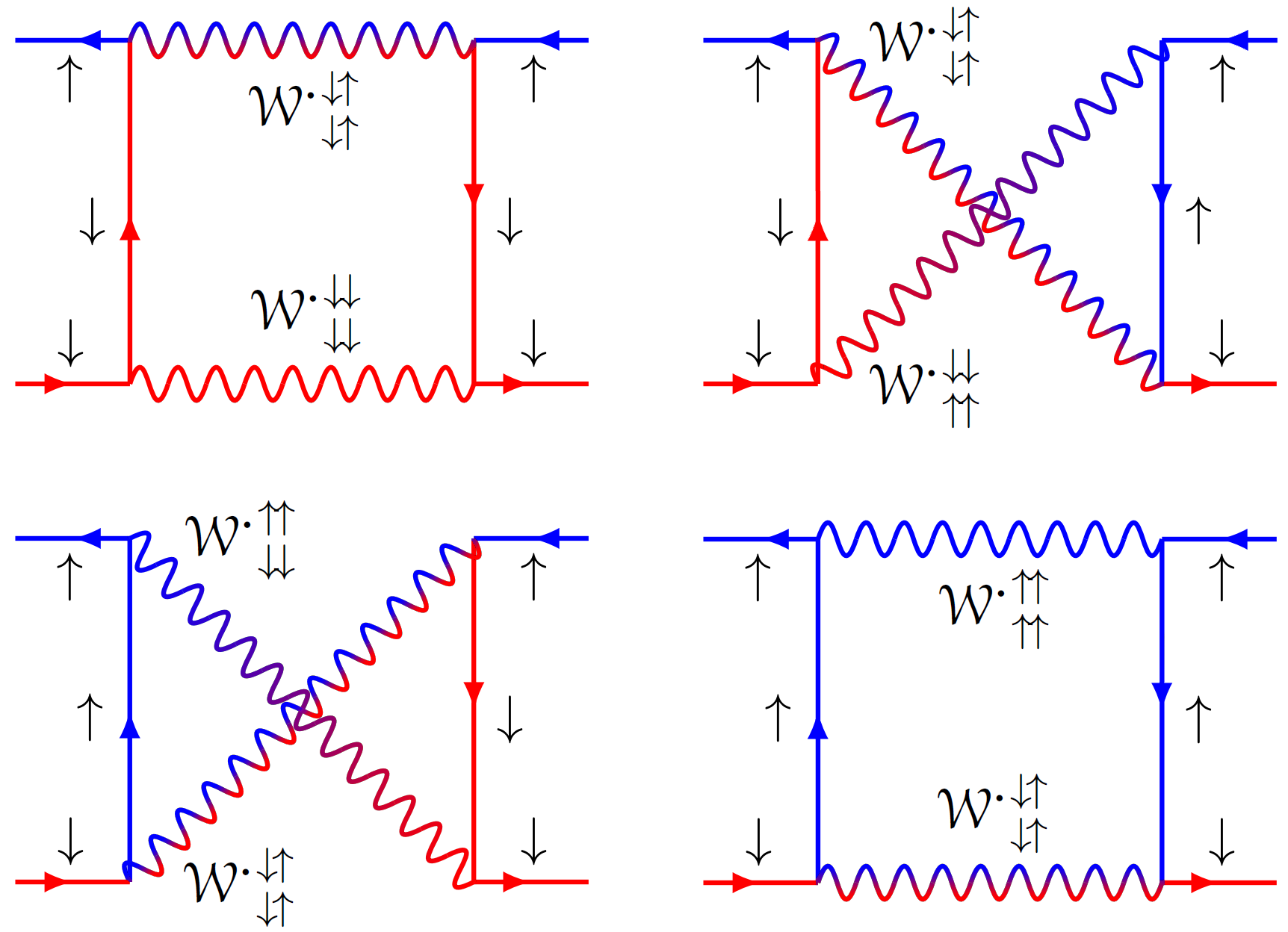}
\caption{Diagrammatic representation of the part of the spin-flip interaction describing simultaneous propagation of spin- and charge-fluctuations. Spin-$\uparrow$ and -$\downarrow$ Green's functions are distinguished by their colors, and spin-conservation in ${\mathcal{W}^{\bcdot }}^{\downarrow \! \uparrow}_{\downarrow \! \uparrow}$ is emphasized with the same colors.} 
\label{fig:fig1}
\end{figure}
\begin{align}     \label{najsest0}
  & {\mathcal{V}^{\bcdot \bcdot}}^{\downarrow \! \uparrow}_{\downarrow \! \uparrow} \lp \mathit{12} ^{\!} , ^{\!} \mathit{34} \rp        =   -  \delta_{\mathit{13}}^{}\delta_{\mathit{24}}^{}  \!
     {\mathcal{W}^{\bcdot}}^{\downarrow \! \downarrow}_{\uparrow \! \uparrow} \lp \mathit{12} \rp    
    \\ 
 &  -       \mathcal{P}^{\downarrow \! \downarrow}_{\downarrow \! \downarrow} \lp \mathit{13} ^{\!} , ^{\!} \mathit{24} \rp 
 \big(   {\mathcal{W}^{\bcdot}}^{\downarrow \! \downarrow}_{\downarrow \! \downarrow} \lp \mathit{13} \rp  {\mathcal{W}^{\bcdot}}^{\downarrow \! \uparrow}_{\downarrow \! \uparrow} \lp \mathit{24} \rp  
  -   \cancel{ {\mathcal{V}^{\bcdot}}^{\downarrow \! \downarrow}_{\downarrow \! \downarrow} \lp \mathit{13} \rp} {\mathcal{V}^{\bcdot}}^{\downarrow \! \uparrow}_{\downarrow \! \uparrow} \lp \mathit{24} \rp 
 \big)    \nonumber 
     \\[-4pt]
  &     -   \mathcal{P}^{\uparrow \! \uparrow}_{\uparrow \! \uparrow} \lp \mathit{13} ^{\!} , ^{\!} \mathit{24} \rp 
  \big(
    {\mathcal{W}^{\bcdot}}^{\downarrow \! \uparrow}_{\downarrow \! \uparrow} \lp \mathit{13} \rp  {\mathcal{W}^{\bcdot}}^{\uparrow \! \uparrow}_{\uparrow \! \uparrow} \lp \mathit{24} \rp     
    -  
    {\mathcal{V}^{\bcdot}}^{\downarrow \! \uparrow}_{\downarrow \! \uparrow} \lp \mathit{13} \rp  \cancel{  {\mathcal{V}^{\bcdot}}^{\uparrow \! \uparrow}_{\uparrow \! \uparrow} \lp \mathit{24} \rp  } 
    \big)  \nonumber   
      \\[-2pt]
  &
  -       \mathcal{P}^{ \downarrow \! \uparrow }_{ \downarrow \! \uparrow } \lp\mathit{13} ^{\!} , ^{\!} \mathit{24}\rp  \big(
   {\mathcal{W}^{\bcdot}}^{\downarrow \! \downarrow}_{\uparrow \!  \uparrow} \lp \mathit{14} \rp  
  {\mathcal{W}^{\bcdot}}^{\downarrow \! \uparrow}_{\downarrow  \! \uparrow } \lp \mathit{23} \rp -  {\mathcal{V}^{\bcdot}}^{\downarrow \! \downarrow}_{\uparrow \!  \uparrow} \lp \mathit{14} \rp  
  {\mathcal{V}^{\bcdot}}^{\downarrow \! \uparrow}_{\downarrow  \! \uparrow } \lp \mathit{23} \rp \big)
      \nonumber 
     \\[-2pt]
  &
    -      \mathcal{P}^{ \uparrow \! \downarrow }_{ \uparrow \! \downarrow } \lp\mathit{13} ^{\!} , ^{\!} \mathit{24}\rp 
  \big(   {\mathcal{W}^{\bcdot}}^{\downarrow \! \uparrow}_{\downarrow \!  \uparrow} \lp \mathit{14} \rp  {\mathcal{W}^{\bcdot}}^{\uparrow \! \uparrow}_{\downarrow \! \downarrow} \lp \mathit{23} \rp    - 
 {\mathcal{V}^{\bcdot}}^{\downarrow \! \uparrow}_{\downarrow \!  \uparrow} \lp \mathit{14} \rp  {\mathcal{V}^{\bcdot}}^{\uparrow \! \uparrow}_{\downarrow \! \downarrow} \lp \mathit{23} \rp   
  \big)
      \nonumber ,
\end{align}   
if ``particle-particle'' two-point contractions of $\mathcal{P}$, with $\mathit{1}=\mathit{4}$ and $\mathit{2}=\mathit{3}$, are neglected. The crossed out spin-components of $\mathcal{V}^{\bcdot}$ in the second and third row are zero due to crossing symmetry, but are kept to illustrate the analogy with the non-vanishing terms in the fourth and fifth row. The four terms that contain two $\mathcal{W}^{\bcdot}$ are presented in Fig. \ref{fig:fig1} and describe the simultaneous propagation of spin and charge excitations. When including lattice vibrations, we will later show that they contain the non-relativistic magnon number-conserving magnon-phonon coupling. If Hedin's formalism was used, the analogous terms to second order in $W$ would miss out on this coupling, since $W$ has no spin-flip components. This shows why the magnon-phonon coupling should be expanded in the collective screened four-point interaction $\mathcal{W}$, which is crossing symmetric, rather than in $W$. Also the first ``screened T matrix'' term \cite{scrT} in Eq. \eqref{najsest0} can be generalized to contain phonons, but their effects in this term are averaged out when treating the spin fluctuations as quasiparticles, as clarified in the following subsection.
\subsection{Two-point approximation in a one-band model} \label{twopoint}
In order to arrive at an efficient approximation with a closer connection to local spin models \cite{Jensen,Cheng,Streib2019,Hellsvik} it is necessary to contract $\mathit{1}$ and $\mathit{2}$ as well as $\mathit{3}$ and $\mathit{4}$ in Eq. \eqref{najsest0}. It is clear from Fig. \ref{fig:fig1} that this leaves intact the simultaneous propagation of spin and charge fluctuations. We assume a one-band model and drop the band index, since interband couplings are of secondary interest to understand the magnon-phonon coupling we aim to find. Dropping also the $\bcdot$'s for convenience, the first term in Eq. \eqref{najsest0} is easily made into a one-point quantity through the replacement 
\begin{align} \label{Wloc}
 & \hspace{0.1cm} \mathcal{W}^{\downarrow \! \downarrow}_{\uparrow \! \uparrow}(\mathit{12})   \coloneqq  \delta_{\mathit{12}}^{}   (\mathcal{U} + \overline{\mathcal{W}}_c ),
 \\ 
 \label{Wcool}
 & \overline{\mathcal{W}}_c^{}    = \int_{-\beta/2}^{\beta/2}  d\tau  \big\langle {\mathcal{W}_c}^{\downarrow \! \downarrow}_{\uparrow \! \uparrow} (\tau)  \big\rangle_{ \! \!  \! \mathsmaller{   \text{ BZ }  } }     .
\end{align}
Here, $\langle \dots \rangle_{ \!   \! \mathsmaller{   \text{ BZ }  } }$ denotes the Brillouin zone average, and we have used that the two-point $\mathcal{W}_c$ only depends on the relative imaginary-time $\tau$. Since the $\tau$-dependence is $\beta$-periodic, the integral is restricted to $[-\beta/2,\beta/2]$ rather than to $[-\beta, \beta]$. The shifted corrections to Eq. \eqref{Wloc} are obtained by replacing $\delta_{\tau_\mathit{1} \tau_\mathit{2}}^{}$, implicit in $\delta_{\mathit{12}}^{}$, with $\delta_{\tau_\mathit{1},\tau_2 \pm \beta}^{}$. These are excluded since they have no effect in Eq. \eqref{eqgreat}, where all imaginary-time integrals are restricted to the interval $[0,\beta]$. Similarly, the free electron-hole pair propagator, which enters into the remaining terms in Eq. \eqref{najsest0}, is replaced by the two-point quantity
\begin{align} \label{conpol}
& \mathcal{P}^{\sigma \sigma'}_{\sigma \sigma'}(\mathit{13} , \mathit{24} )    \coloneqq \delta_{\mathit{12}}^{} \delta_{\mathit{34}}^{} \overline{\CMcal{G}}_\sigma^{} \overline{\CMcal{G}}_{\sigma'}^{},
 \\
  \label{locG}
& \hspace{0.5cm} \overline{\CMcal{G}}_\sigma^{}  =  \int_{-\beta/2}^{\beta/2}  d\tau  \big\langle  \CMcal{G}_\sigma^{} (\tau)  \big\rangle_{ \! \!  \! \mathsmaller{   \text{ BZ }  } }   .
\end{align}
This integral would vanish if it ranged from $-\beta$ to $\beta$, due to the antisymmetry of $\CMcal{G}_\sigma^{}$ under shifts of $- \beta$ in the imaginary-time interval $\tau \in [0,\beta]$. But despite this antisymmetry, the contributions outside the interval $[-\beta/2,\beta/2]$ are suppressed in Eq. \eqref{najsest0}, motivating the choice of Eq. \eqref{locG}. It is of interest to find an expression for $\overline{\CMcal{G}}_\sigma^{}$ since it later will be shown to appear in the magnon-phonon coupling when including lattice dynamics. Since we have already broken self-consistency in Sec. \ref{secit}, we assume that the Green's function is obtained from a single-particle calculation. In practice this, may be a Hartree-Fock or a (spin-)density functional theory \cite{jacob} calculation. We can then write
\begin{align} \label{ost}
  \big\langle \CMcal{G}_\sigma^{} \lp \tau \rp \big\rangle_{ \! \!  \! \mathsmaller{   \text{ BZ }  } }  & \! =  \Big\langle   \frac{ e^{-\xi_\sigma^{}  \tau }   }{e^{\xi_\sigma^{}  \beta } \! + \! 1  } 
 \Big\rangle_{ \! \!  \! \mathsmaller{   \text{ BZ }  } }  \! \! \theta ( \!-\tau \rp  -  \Big\langle  \frac{e^{\xi_\sigma^{}  (\beta-\tau) }      }{e^{\xi_\sigma^{} \beta } \! + \!  1  }
 \Big\rangle_{ \! \!  \! \mathsmaller{   \text{ BZ }  } }  \! \!    \theta \lp \tau \rp , 
\end{align}  
$\xi_\sigma^{} $ is shorthand notation for the momentum dependent spin-$\sigma$ electronic dispersion, $\xi_\sigma^{}({\boldsymbol k})$, measured relative to the chemical potential $\mu$. The two terms are strongly peaked close to $\tau=0$ if $\xi_\sigma^{}  \beta \ll 0$ and $\xi_\sigma^{}  \beta \gg 0$, respectively, which approximately holds if $\xi_\sigma^{} < 0$ and $\xi_\sigma^{} > 0$, since typical electronic energy scales are large compared to thermal energies. Consistent with this is the replacement of the exponential functions in the two terms of Eq. \eqref{ost} with properly normalized delta functions,
\begin{align} \label{ost2}
&  \big\langle \CMcal{G}_\sigma^{} ( \tau ) \big\rangle_{ \! \!  \! \mathsmaller{   \text{ BZ }  } }   \! \approx \tfrac{1}{2}  \Big\langle   \frac{  \theta(-\xi_\sigma^{} )  }{e^{\xi_\sigma^{}  \beta } \! + \! 1  } 
  \int_{-\beta}^0 d\tau'
 e^{-\xi_\sigma^{}  \tau' }   
 \Big\rangle_{ \! \!  \! \mathsmaller{   \text{ BZ }  } }  \!   \delta(\tau)  \\[-2pt]
 &  \hspace{1.4cm} - \tfrac{1}{2}  \Big\langle  \frac{e^{\xi_\sigma^{}  \beta }    \theta(\xi_\sigma^{})  }{e^{\xi_\sigma^{} \beta } \! + \!  1  }  \int_0^{\beta} d\tau' e^{-\xi_\sigma^{}  \tau' }
 \Big\rangle_{ \! \!  \! \mathsmaller{   \text{ BZ }  } }  \!   \delta(\tau)   , \nonumber  
\end{align}  
where $\theta(0)=\tfrac{1}{2}$ has been used twice. By integrating Eq. \eqref{ost2} and using the Fermi occupations $n_\sigma^{} = n_F^\beta(\xi_\sigma^{})$, where $n_F^\beta(\omega)  = (e^{\omega \beta} +1 )^{-1}$, Eq. \eqref{locG} becomes
\begin{align} \label{finalG}
\overline{\CMcal{G}}_{\sigma}^{} & =  \Big\langle  \frac{    n_\sigma^{ }  - \mathsmaller{1/2}}{ |\xi_\sigma^{}| } 
 \Big\rangle_{ \! \!  \! \mathsmaller{   \text{ BZ }  } }  \! .
\end{align}
This does in general not vanish at half-filling. The divergence of $1/|\xi_\sigma^{}|$ cancels by the numerator. The two-point approximation to the spin-flip interaction in Eq. \eqref{najsest0} can now be expressed as 
\begin{align}   \label{najsestnew}
   {\mathcal{V}}^{\downarrow \! \uparrow}_{\downarrow \! \uparrow} \lp \mathit{12} ^{\!} , ^{\!} \mathit{34} \rp  &  = - \delta_{\mathit{12}}^{} \delta_{\mathit{23}}^{} \delta_{\mathit{34}}^{} \mathcal{U}  + \delta_{\mathit{12}}^{} \delta_{\mathit{34}}^{} \Delta^{\downarrow \! \uparrow}_{\downarrow \! \uparrow} \lp \mathit{13}  \rp  ,
\end{align} 
where Eqs. \eqref{Wloc} and \eqref{conpol} can be used to show that
\begin{align} \label{najsestnew2}
& \hspace{-0.1cm} \Delta^{\downarrow \! \uparrow}_{\downarrow \! \uparrow} \lp \mathit{12}  \rp   =     - \delta_{\mathit{12}}^{} \mathcal{U}'  \! - \! \!  \sum_{\sigma_\mathit{1}   \sigma_\mathit{2}}  \! \overline{\CMcal{G}}_{\sigma_\mathit{1}}^{ }   \overline{\CMcal{G}}_{\sigma_\mathit{2}}^{ } \!  {\mathcal{W}_c^{}}^{\downarrow \! \uparrow}_{\downarrow \! \uparrow} \lp \mathit{12} \rp    
   { \mathcal{W}_c^{}}^{\sigma_\mathit{1} \sigma_\mathit{1}}_{\sigma_\mathit{2} \sigma_\mathit{2}} \lp \mathit{12} \rp    ,    \!   
 \\ 
 \label{secondd}
 & \mathcal{U}'   =  \mathcal{U} \!  \sum_{\sigma_\mathit{1}     \sigma_\mathit{2}}  \! \overline{\CMcal{G}}_{\sigma_\mathit{1}}^{ }   \overline{\CMcal{G}}_{\sigma_\mathit{2}}^{ }  \! \big( ^{\!} \sigma^x_{\sigma_\mathit{1}  \sigma_\mathit{2}}   \!  {\mathcal{W}_c^{}}^{\downarrow \! \uparrow}_{\downarrow \! \uparrow} \lp \mathit{1}   \rp  \!-\!   {\mathcal{W}_c^{}}^{\sigma_\mathit{1} \sigma_\mathit{1}}_{\sigma_\mathit{2} \sigma_\mathit{2}} \lp \mathit{1}   \rp ^{\!}   \big)     \! + \!  \overline{\mathcal{W}}_c^{}   , \!
\end{align}
where we have abandoned Einstein's summation convention. $\mathcal{U}'$ is a correction to the local interaction, and $\mathit{1}$ in Eq. \eqref{secondd} is short for $\mathit{11}$. The two-point structure in Eq. \eqref{najsestnew2} allows for an efficient calculation of the spin susceptibility, since Eq. \eqref{eqgreat} --- which has four-point structure, can be replaced by the two-point equations 
\begin{align} \label{eqgreatnew}
    \mathcal{R}^{\downarrow \! \uparrow}_{\downarrow \! \uparrow}\lp \mathit{12} \rp  &   =  r^{\downarrow \! \uparrow}_{\downarrow \! \uparrow} \lp \mathit{12} \rp  + \!  \int  \! d (\mathit{34})   r^{\downarrow \! \uparrow}_{\downarrow \! \uparrow}  \lp \mathit{13} \rp \Delta^{\downarrow \! \uparrow}_{\downarrow \! \uparrow} \lp \mathit{34} \rp  \mathcal{R}^{\downarrow \! \uparrow}_{\downarrow \! \uparrow} \lp \mathit{42} \rp   , \\
    \label{littler}
     r^{\downarrow \! \uparrow}_{\downarrow \! \uparrow}\lp \mathit{12} \rp  &   =  \mathcal{P}^{\downarrow \! \uparrow}_{\downarrow \! \uparrow} \lp \mathit{12} \rp  -  \mathcal{U} \int  \! d \mathit{3}   \mathcal{P}^{\downarrow \! \uparrow}_{\downarrow \! \uparrow}  \lp \mathit{13} \rp     r^{\downarrow \! \uparrow}_{\downarrow \! \uparrow} \lp \mathit{32} \rp ,
\end{align}
where the two-point $\mathcal{R}$ is defined like in Eq. \eqref{susc}, and $r$ is the unperturbed spin susceptibility obtained from the Fock exchange only. It will in the next subsection enter as the magnons in the magnon-phonon coupling. In momentum and Matsubara space, the {\it renormalized} magnons of Eq. \eqref{eqgreatnew} can be written in terms of the Fock magnons as 
\begin{align} \label{eqgreatnew2}
    \mathcal{R}^{\downarrow \! \uparrow}_{\downarrow \! \uparrow}\lp  {\boldsymbol k}, i\omega_m \rp  &   =  \frac{  r^{\downarrow \! \uparrow}_{\downarrow \! \uparrow} \lp  {\boldsymbol k}, i \omega_m \rp }{1 - r^{\downarrow \! \uparrow}_{\downarrow \! \uparrow} \lp  {\boldsymbol k}, i \omega_m \rp \Delta^{\downarrow \! \uparrow}_{\downarrow \! \uparrow}\lp  {\boldsymbol k}, i \omega_m \rp }   ,
\end{align}
where $\omega_m = 2\pi m/\beta$. From Eq. \eqref{najsestnew2} it follows that 
\begin{align} \label{najsestnew3}
  &  \hspace{3cm}  \Delta^{\downarrow \! \uparrow}_{\downarrow \! \uparrow} (  {\boldsymbol k}, i \omega_m )   =       -\mathcal{U}'  \\
   &    -  \! \frac{1}{N \beta} \sum_{\substack{{\boldsymbol q} \omega_n \\ \sigma_\mathit{1} \sigma_\mathit{2} }}  \! \overline{\CMcal{G}}_{\sigma_\mathit{1}}^{ }   \overline{\CMcal{G}}_{\sigma_\mathit{2}}^{ } \!  {\mathcal{W}_c^{}}^{\downarrow \! \uparrow}_{\downarrow \! \uparrow} (   {\boldsymbol k} \!- \! {\boldsymbol q},  i \omega_m \!-\! i \omega_n )  
    {\mathcal{W}_c^{}}^{\sigma_\mathit{1} \sigma_\mathit{1}}_{\sigma_\mathit{2} \sigma_\mathit{2}} (  {\boldsymbol q},  i \omega_n ) \nonumber  ,
\end{align} 
where $N$ is the number of ${\boldsymbol k}$ points in the Brillouin zone.
\subsection{Magnon-phonon coupling}
We now deform the $^{\sigma_\mathit{1} \sigma_\mathit{1}}_{\sigma_\mathit{2} \sigma_\mathit{2}}$-component of $\mathcal{W}_c^{}$ in Eq. \eqref{najsestnew3} to include phonons, as anticipated throughout this article. Following the works of Hedin \cite{hedin69} and Giustino \cite{Giustino2017}, where the nuclear spin is neglected, it can be split into a term of electronic charge fluctuations (plasmons) and one which contains the phonons. Assuming a single phonon mode $ \nu = \mathsmaller{\text{P}}$ ($\mathsmaller{\text{P}}$ for phonon), we get  
\begin{align} \label{important}
& \hspace{1.9cm}  {\mathcal{W}_c^{}}^{\sigma_\mathit{1} \sigma_\mathit{1} \!}_{\sigma_\mathit{2} \sigma_\mathit{2}\!} (  {\boldsymbol q},^{\!}  i \omega_{n} ) \nonumber      \\
 & \coloneqq  {\mathcal{W}_c^{}}^{\sigma_\mathit{1} \sigma_\mathit{1}\!}_{\sigma_\mathit{2} \sigma_\mathit{2}\!} (  {\boldsymbol q},^{\!}  i \omega_{n} )  
   +   g_{  {\boldsymbol q} }^{\sigma_\mathit{1}  \! }   
g_{{\boldsymbol q} }^{  \sigma_{\mathit{2}}  * }   \mathcal{D} ( {\boldsymbol q},^{\!} i \omega_{n} )    , 
\end{align}
where   
\begin{align} \label{eqcool2}
 &  g_{{\boldsymbol q} }^{\sigma}    = \frac{1}{N} \sum_{{\boldsymbol k}} g_{nn,\nu = \mathsmaller{\text{P}} \! }^\sigma({\boldsymbol k},{\boldsymbol q})  
 \end{align} 
is the one-momentum electron-phonon coupling between the electrons in the band of our one-band model, $n$ (which we drop), and the phonons in branch $ \nu = \mathsmaller{\text{P}}$, and 
\begin{align} \label{phonon}
\mathcal{D}({\boldsymbol q},i\omega_n) & = \frac{2 \omega_{{\boldsymbol q}}^{{\mathsmaller{\text{P}}}}}{(i \omega_n)^2  - ( \omega_{{\boldsymbol q}}^{{\mathsmaller{\text{P}}}} )^2}
\end{align}
is the phonon propagator in the adiabatic approximation, determined from the phonon dispersion $\omega_{{\boldsymbol q}}^{{\mathsmaller{\text{P}}}} \geq 0$, neglecting life-time broadening. Nothing in our formalism requires this simplification, but it allows for analytic Matsubara summation. Extending to several electron bands and phonon modes is straightforward, but the momentum average and band diagonality in Eq. \eqref{eqcool2} follow from the two-point contraction in \ref{twopoint}. From Ref. \cite{Giustino2017}, we find
 \begin{align} \label{neat}
g_{ {\boldsymbol q} }^{ \sigma}   & =  \frac{ -     Z    }{\sqrt{2 M  \omega_{{\boldsymbol q}}^{{\mathsmaller{\text{P}}}}   }}      \big( {\boldsymbol e}_{{\boldsymbol q}}^{{\mathsmaller{\text{P}}}}  \cdot      {\boldsymbol {\mathcal F}}_{ {\boldsymbol q} }^{ \sigma }  \big) 
\end{align} 
if assuming a single (light) atom in each unit cell, with atomic number $Z$, mass $M$ and equilibrium position ${\boldsymbol \tau}^{(0)}_{}$ in the central cell. ${\boldsymbol e}_{{\boldsymbol q}}^{{\mathsmaller{\text{P}}}} $ is the phonon polarization, and 
\begin{align}
{\boldsymbol {\mathcal F}}_{ {\boldsymbol q}}^{ \sigma}    & =  \sum_{   {\boldsymbol T}  } e^{i {\boldsymbol q} \cdot {\boldsymbol T}}     \int d{\boldsymbol r}   w_{ {\boldsymbol 0} }^{ *}({\boldsymbol r})           
{\boldsymbol {\mathcal F}}^{}_{\! \sigma} \! ({\boldsymbol r} ,    {\boldsymbol T}) 
w_{ {\boldsymbol 0} }({\boldsymbol r}) 
\end{align} 
are Fourier components of the diagonal Wannier matrix elements in the central unit cell of the ``screened force'' from the nuclei (or ions) in the ${\boldsymbol T}$-shifted unit cells,
\begin{align}
{\boldsymbol {\mathcal F}}^{}_{ \! \sigma \! } ({\boldsymbol r} ,  {\boldsymbol T}) & =   \sum_{\sigma'} \!  \int  d{\boldsymbol r}'
 [\mathcal{S}^{\mathsmaller{-1}} ]^{\sigma \sigma}_{\sigma' \sigma'} ({\boldsymbol r} {\boldsymbol r}')       
  \frac{   ( {\boldsymbol r}' \! - \!  {\boldsymbol T}   \! - \!   {\boldsymbol \tau}_{ }^{(0)} ) }{|{\boldsymbol r}'  \! - \!  {\boldsymbol T}   \!  -  \!  {\boldsymbol \tau}_{  }^{(0)}|^3} .
\end{align}  
$\mathcal{S}^{\mathsmaller{-1}}$ is here the spin-dependent finite-temperature analog of the static $\epsilon^{-1}$ in Ref. \cite{Giustino2017}. Equation \eqref{important} and ($\bar{\sigma}=-\sigma$) 
\begin{align}
{\mathcal{W}_c}^{\downarrow \! \uparrow}_{\downarrow \! \uparrow} ({\boldsymbol q},i\omega_n)   & =  \mathcal{U}^2
{r}^{\downarrow \! \uparrow}_{\downarrow \! \uparrow} ({\boldsymbol q},i\omega_n) 
 , \\
 {\mathcal{W}_c}^{\sigma_\mathit{1} \sigma_\mathit{1}}_{\sigma_{\mathit{2}} \sigma_{\mathit{2}}} ({\boldsymbol q},i\omega_n)   & = \mathcal{U}^2
{r}^{\bar{\sigma}_\mathit{1} \bar{\sigma}_\mathit{1} }_{ \bar{\sigma}_\mathit{2} \bar{\sigma}_\mathit{2}} ({\boldsymbol q},i\omega_n) 
 ,
\end{align}
which hold since $\mathcal{W}_c = \mathcal{W}_c^{\bcdot}$, turn Eq. \eqref{najsestnew3} into 
\begin{align} \label{najsestnew4}
  &  \Delta^{\downarrow \! \uparrow}_{\downarrow \! \uparrow} (  {\boldsymbol k}, i \omega_m )   =      - \mathcal{U}'_\mathcal{D}    \! -  \!   \frac{\mathcal{U}^2}{N \beta} \! \sum_{\substack{{\boldsymbol q} \omega_n \\ \sigma_\mathit{1} \sigma_\mathit{2} }} \overline{\CMcal{G}}_{\sigma_\mathit{1}}^{ }   \overline{\CMcal{G}}_{\sigma_\mathit{2}}^{ } r^{\downarrow \! \uparrow}_{\downarrow \! \uparrow} (   {\boldsymbol k} \!- \! {\boldsymbol q},  i \omega_m \!-\! i \omega_n )  \nonumber  \\ 
   &  \hspace{1.15cm} \smalltimes  \! \big( \mathcal{U}^2   
   r^{\bar{\sigma}_\mathit{1} \bar{\sigma}_\mathit{1}}_{\bar{\sigma}_\mathit{2} \bar{\sigma}_\mathit{2}} (  {\boldsymbol q},  i \omega_n )    \!+ \!   
   g_{ {\boldsymbol q} }^{\sigma_\mathit{1}}  
 \mathcal{D} ( {\boldsymbol q},i \omega_n ) g_{ {\boldsymbol q} }^{ \sigma_\mathit{2} * }    \big)     ,
\end{align}
where $\mathcal{U}'_\mathcal{D}$ is the modified $\mathcal{U}'$ due to the phonons. The first term in the parenthesis describes the coupling between Fock magnons and longitudinal spin fluctuations as well as electronic charge fluctuations (plasmons), and the second term the magnon-phonon coupling. Since, typically, only the magnon and phonon energies are comparable, we keep only the last term in the following. Furthermore, the term $\mathcal{U}'_\mathcal{D}$ will be considered fixed by the Goldstone criterion, which requires that the magnon dispersion contained in the imaginary part of $\mathcal{R}^{\downarrow \! \uparrow}_{\downarrow \! \uparrow}$ in Eq. \eqref{eqgreatnew2} approaches zero in the long-wavelength limit, ${\boldsymbol k} \to {\boldsymbol 0}$. It remains to find $r^{\downarrow \! \uparrow}_{\downarrow \! \uparrow}$, but since it is generated by the local and instantaneous interaction $\mathcal{U}$, it lacks lifetime broadening and is accurately parametrized as  
\begin{align} \label{magnon}
r^{\downarrow \! \uparrow}_{\downarrow \! \uparrow}({\boldsymbol k}, i \omega_m) & =  \frac{1}{i \omega_m  -    \omega_{{\boldsymbol k}}^{{\mathsmaller{\text{M}}}} } ,
\end{align}
in term of the Fock magnon dispersion $ \omega_{{\boldsymbol k}}^{{\mathsmaller{\text{M}}}} \geq 0$, treated as temperature-independent for simplicity. The positivity holds since $\uparrow$ is the majority spin channel. In practice, Eq. \eqref{magnon} can be matched to first-principle calculations of $r^{\downarrow \! \uparrow}_{\downarrow \! \uparrow}$ based on Eq. \eqref{littler}. The magnon-phonon coupling in Eq. \eqref{najsestnew4} can be written on a physically transparent form by making use of Eqs. \eqref{phonon} and \eqref{magnon} and performing Matsubara summation and analytic continuation ($i\omega_m \to  \omega + i \eta$). The result is 
\begin{align} \label{najsestnew5}
  & \hspace{0.93cm}  {\Delta_{\mathsmaller{\text{MP}}}^{r}}^{\downarrow \! \uparrow}_{\downarrow \! \uparrow} (  {\boldsymbol k}, \omega )   =    \mathcal{U}^2  \!  \int \!   \frac{d{\boldsymbol q}}{ \Omega_{\mathsmaller{\text{BZ}}}^{} }  \big|   \overline{\CMcal{G}}_{\uparrow }^{ }  g_{ {\boldsymbol q}}^{\uparrow}  + \overline{\CMcal{G}}_{\downarrow }^{ }  g_{{\boldsymbol q}}^{\downarrow}       \big|^2     \\ 
   &   \smalltimes \!   \bigg(    \frac{
n_{{\boldsymbol q}}^{{\mathsmaller{\text{P}}}}
 -  n_{{\boldsymbol k}-{\boldsymbol q}}^{{\mathsmaller{\text{M}}}}  }{ \omega  +   
\omega_{{\boldsymbol q}}^{{\mathsmaller{\text{P}}}} 
 -    
 \omega_{{\boldsymbol k}-{\boldsymbol q}}^{{\mathsmaller{\text{M}}}}  
 +  i \eta   }   +  \frac{ 1  +  n_{{\boldsymbol q}}^{{\mathsmaller{\text{P}}}}   +     n_{{\boldsymbol k}-{\boldsymbol q}}^{{\mathsmaller{\text{M}}}}  }{ \omega   -  \omega_{{\boldsymbol q}}^{{\mathsmaller{\text{P}}}}   -   \omega_{{\boldsymbol k}-{\boldsymbol q}}^{{\mathsmaller{\text{M}}}}   +  i \eta }  \bigg)  \nonumber  ,
\end{align}
where $n^{\mathsmaller{\text{M/P}\!}}_{{\boldsymbol q}} =  n_B^\beta( \omega^{\mathsmaller{\text{M/P}\!}}_{{\boldsymbol q}}  ) $ and $n_B^\beta(\omega) = (e^{ \beta  \omega   }-1)^{-1}$ is the magnon/phonon Bose occupation. The superscript $r$ emphasizes that it is the retarded coupling. Likewise, the retarded Fock magnon propagator ${r^r}^{\downarrow \! \uparrow}_{\downarrow \! \uparrow}$ and renormalized magnon propagator ${\mathcal{R}^r}^{\downarrow \! \uparrow}_{\downarrow \! \uparrow}$ are obtained by analytically continuing Eqs. \eqref{magnon} and \eqref{eqgreatnew2}, respectively, and the latter is determined from the former and Eq. \eqref{najsestnew5}, up to the Goldstone shift. In Eq. \eqref{najsestnew5}, the continuous limit $\tfrac{1}{N} \sum_{\boldsymbol k} \! \to \! \int \tfrac{d{\boldsymbol q}}{\Omega_{\mathsmaller{\text{BZ}}}^{}}$ has been taken, where $\Omega_{\mathsmaller{\text{BZ}}}^{}$ is the Brillouin zone volume. The two terms describe phonon absorption and emission, respectively, and only the emission term survives in the limit $T \to 0$. Except from the fact that Eq. \eqref{najsestnew5} is not restricted to acoustic phonons, it has the same form as the non-relativistic contribution to Eq. (4.4) in Ref. \cite{Streib2019}, where the coupling was derived using a phenomenological magnetoelastic model. This suggests that the present work provides a path beyond such models, by avoiding one or several of the assumptions made to arrive at Eq. \eqref{najsestnew5}. The inclusion of spin-orbit coupling, and consequently anisotropy, will be discussed in a future paper.
\section{Model calculations}
\subsection{Introducing the model}
A minimal 3-dimensional model is considered where the magnon-phonon coupling in Eq. \eqref{najsestnew5} is made momentum-independent, while retaining its frequency dependence. To this aim, we assume an isotropic magnon dispersion $\omega^{\mathsmaller{\text{M}}}_{q}$, where $q=|{\boldsymbol q}|$, a dispersion-less optical phonon $\omega^{\mathsmaller{\text{P}}}_{}$, replace the spin-dependent electron-phonon coupling $g^\sigma_{{\boldsymbol q}}$ with its spin-average $g_{{\boldsymbol q}}$, replace $|g_{\boldsymbol q}|^2$ with its Brillouin zone-average $\overline{g}^2$ \cite{Goodvin2008}, 
and finally approximate the Brillouin zone-integral of any isotropic function with an integral over a sphere with matching volume, i.e. with radius $K=(3\Omega_{\mathsmaller{\text{BZ}}}^{} /4\pi)^{1/3}$. The magnon-phonon coupling in Eq. \eqref{najsestnew5} then takes the simple form  
\begin{align} \label{najsestnew6}
  & \hspace{3.23cm}  {\Delta_{\mathsmaller{\text{MP}}}^{r}}^{\downarrow \! \uparrow}_{\downarrow \! \uparrow} ( \omega )   =    \mathcal{A}^2     \\
   &   \smalltimes  \! \int_0^K  \! \frac{4 \pi q^2 dq }{ \Omega_{\mathsmaller{\text{BZ}}}^{} }   \bigg(    \frac{
n^{\mathsmaller{\text{P}}}_{ } 
 \! - n^{\mathsmaller{\text{M}}}_{q}   }{ \omega +  \omega^\mathsmaller{\text{P}}_{}   \! -   \omega^{\mathsmaller{\text{M}}}_{q}   + i \eta   }   +  \frac{ 1  +  n^{\mathsmaller{\text{P}}}_{ }  \!   +   n^{\mathsmaller{\text{M}}}_{q }   }{ \omega  -   \omega^\mathsmaller{\text{P}}_{}  \!  -   \omega^{\mathsmaller{\text{M}}}_{q}  + i \eta }  \bigg)  \nonumber  
     ,
\end{align}
where we have introduced the (positive) magnon-phonon coupling strength 
\begin{align}
 \mathcal{A} & =  \mathcal{U}    \big| \overline{\CMcal{G}}_{\uparrow }^{ } +  \overline{\CMcal{G}}_{\downarrow }^{ } \big| \sqrt{ \overline{g}^2 }  . 
\end{align}
\begin{figure}[h]
\centering
\includegraphics[width=0.82\linewidth]{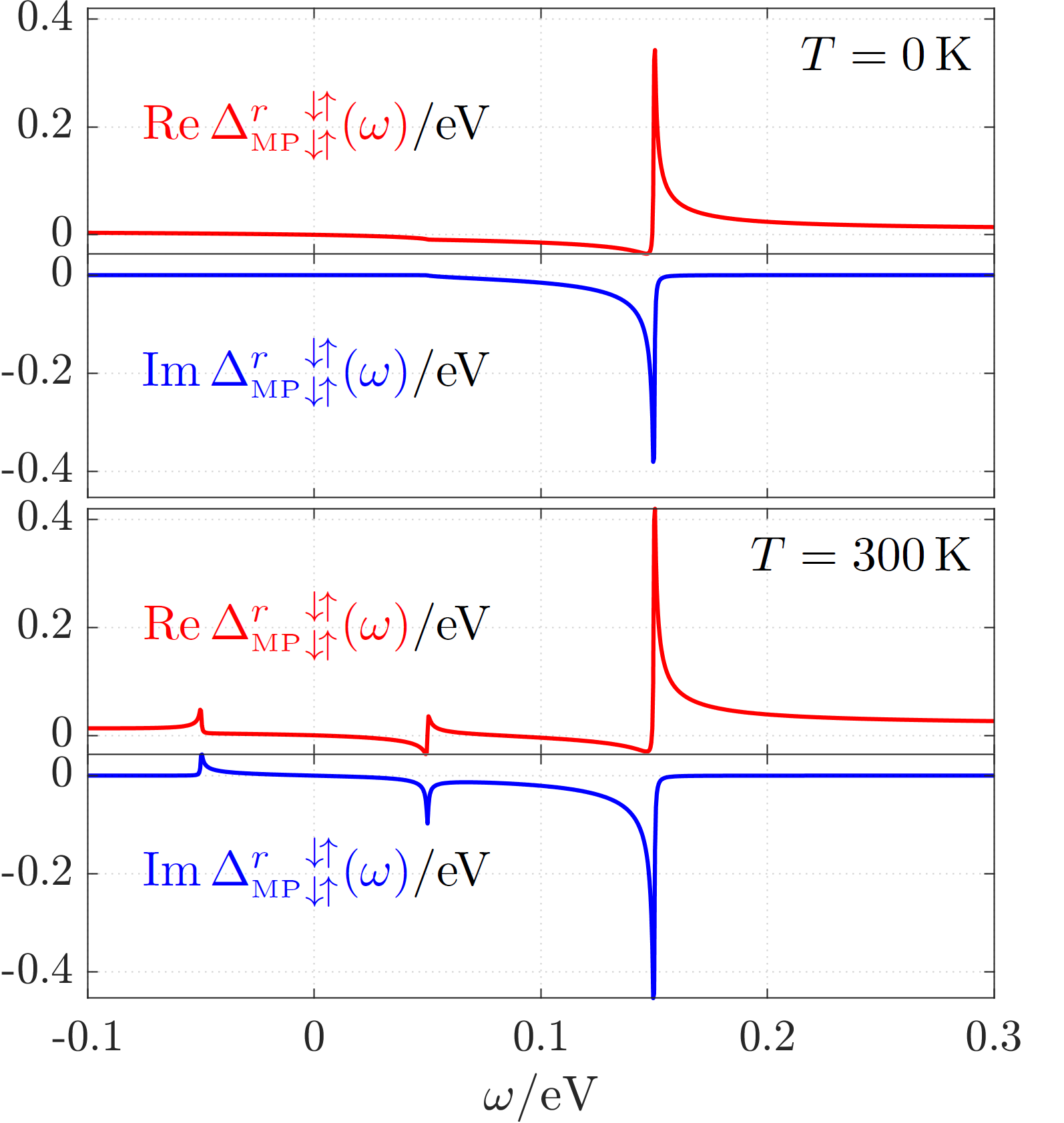}
\caption{Real and imaginary parts of ${\Delta_{\mathsmaller{\text{MP}}}^{r}}^{\downarrow \! \uparrow}_{\downarrow \! \uparrow} ( \omega ) $ at $T=0\operatorname{K}$ and $T=300\operatorname{K}$, with $\mathcal{A}=32 \operatorname{meV}$.} 
\label{fig:fig2}
\end{figure} 
\begin{figure}[h]
\centering
\includegraphics[width=0.97\linewidth]{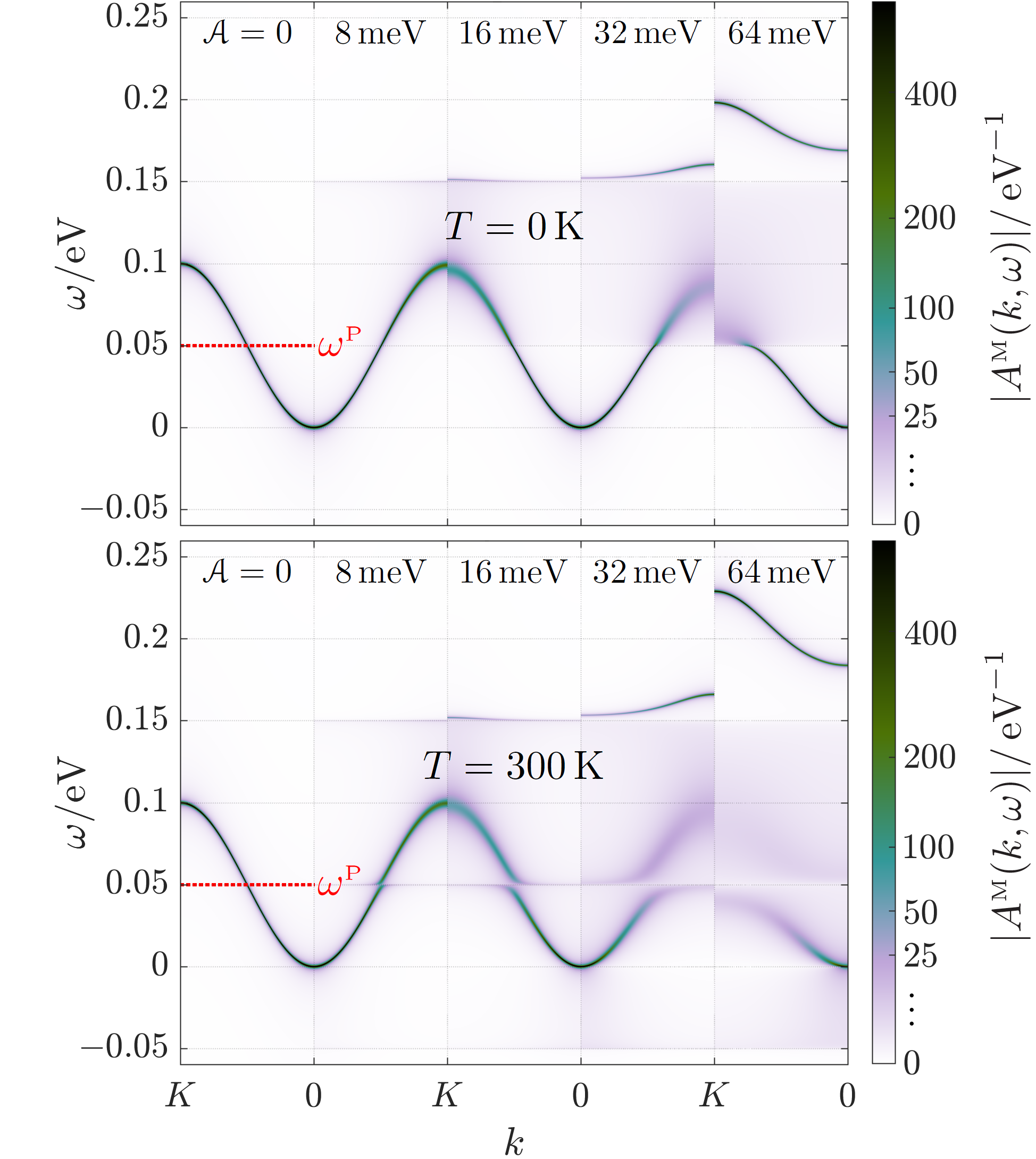}
\caption{Magnititude of the renormalized magnon spectral function ($A^{\mathsmaller{\text{M}}}_{}(k,\omega) $) at $T=0\operatorname{K}$ and $T=300\operatorname{K}$ for various $\mathcal{A}$.} 
\label{fig:fig3}
\end{figure}
\noindent This has the dimension of energy, for which we in the following will use the units of electron volts rather than the Hartree units. In the results that will follow, we tune $\mathcal{A}$ independently of temperature. We also use a typical optical phonon energy of $  \omega^\mathsmaller{\text{P}}_{}    = 0.05 \operatorname{eV}$ chosen to be in the middle of the isotropic magnon band, which we parametrize as $  \omega^{\mathsmaller{\text{M}}}_{q}   = 0.1 \sin^2 (\tfrac{ q \pi}{2 K}) \operatorname{eV}$, with the typical value $K=\pi/a$, with $a=7$ in atomic units, assumed temperature-independent. The magnon-phonon coupling will be complemented with a constant shift to guarantee that the Goldstone criterion is satisfied. 
\subsection{Results}
\begin{figure}[h]
\includegraphics[width=0.97\linewidth]{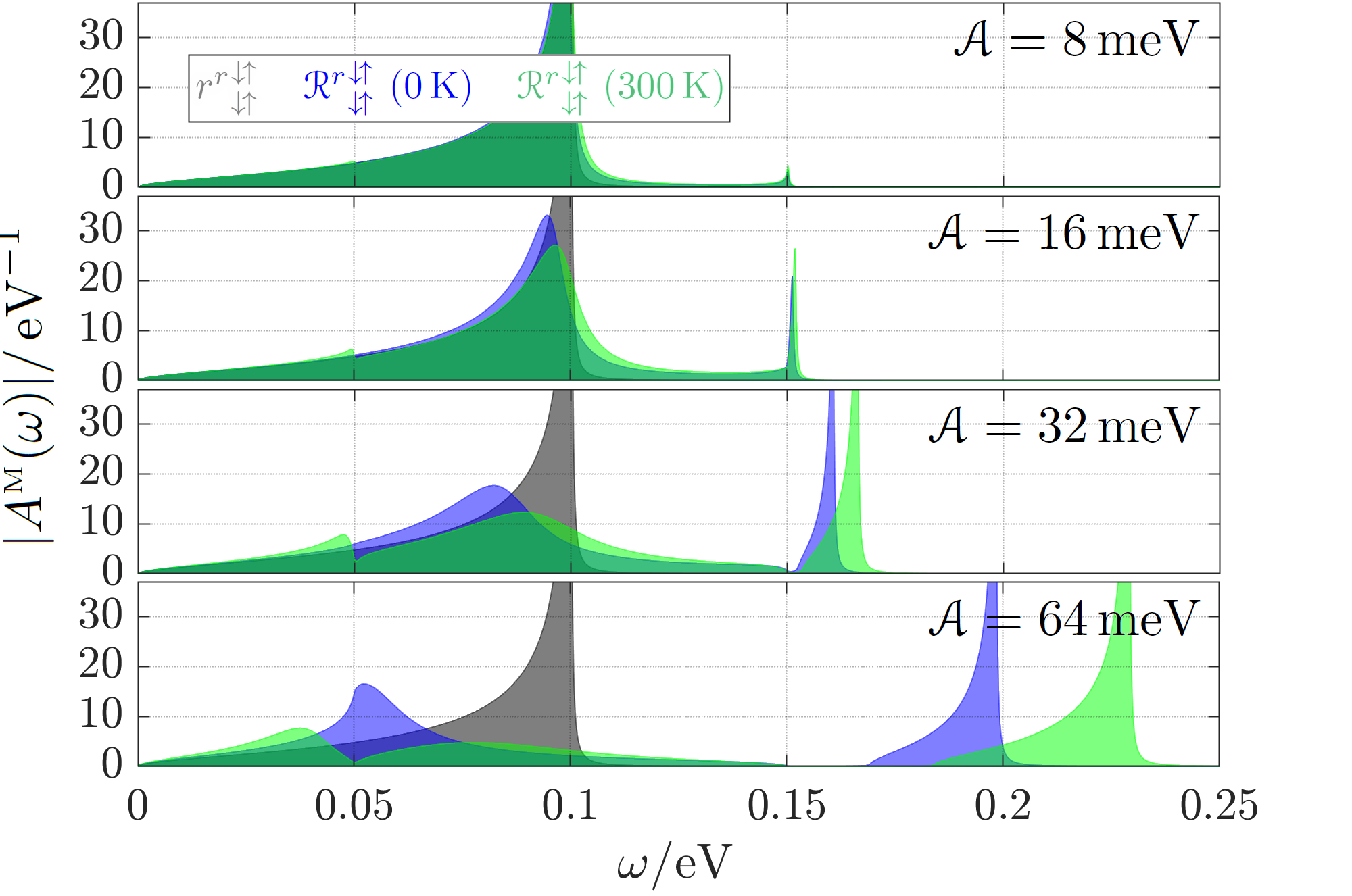}
\caption{Magnitude of the total spectral function, ($A^{\mathsmaller{\text{M}}}_{}(\omega)$), for Fock magnons (${r^r}^{\downarrow \! \uparrow}_{\downarrow \! \uparrow}$) and renormalized magnons (${\mathcal{R}^r}^{\downarrow \! \uparrow}_{\downarrow \! \uparrow}$) at $T=0\operatorname{K}$ and $T=300\operatorname{K}$ for various $\mathcal{A}$.} 
\label{fig:fig4}
\end{figure}
The real and imaginary parts of the retarded magnon-phonon coupling ${\Delta_{\mathsmaller{\text{MP}}}^{r}}$ in Eq. \eqref{najsestnew6} are presented in Fig. \ref{fig:fig2} at temperatures $0\operatorname{K}$ and $300\operatorname{K}$ (room temperature) for a coupling strength of $\mathcal{A}=32 \operatorname{meV}$. The convergence parameter $\eta$ is chosen to be $0.3\operatorname{meV}$. The results for different values of $\mathcal{A}$ are the same up to the overall scaling. The main dispersion feature present both at absolute zero and at room temperature describes phonon emission and occurs around the energy $0.15\operatorname{eV}$. This therefore originates partly from the ``zero-temperature term'' in Eq. \eqref{najsestnew6} that lacks both phonon and Fock magnon (FM) Bose occupation factors. The energy is understood by adding the phonon energy $0.05\operatorname{eV}$ and the energy $0.1\operatorname{eV}$, for which the FM spectral density is the largest. Temperature enhances this feature but also induces dispersion features corresponding to phonon absorption at the energies $\pm 0.05 \operatorname{eV}$, caused by the first term in Eq. \eqref{najsestnew6}. This gets large either for FM energies close to $0$, where the Bose occupation is large, or for FM energies close to $0.01\operatorname{eV}$, with large spectral weight, explaining the two features. In our particular model, the phonon energy of $0.05\operatorname{eV}$ coincides with the energy difference between the predominant FM energy of $0.1\operatorname{eV}$ and the phonon energy.

\begin{figure}[t]
\centering
\includegraphics[width=0.99\linewidth]{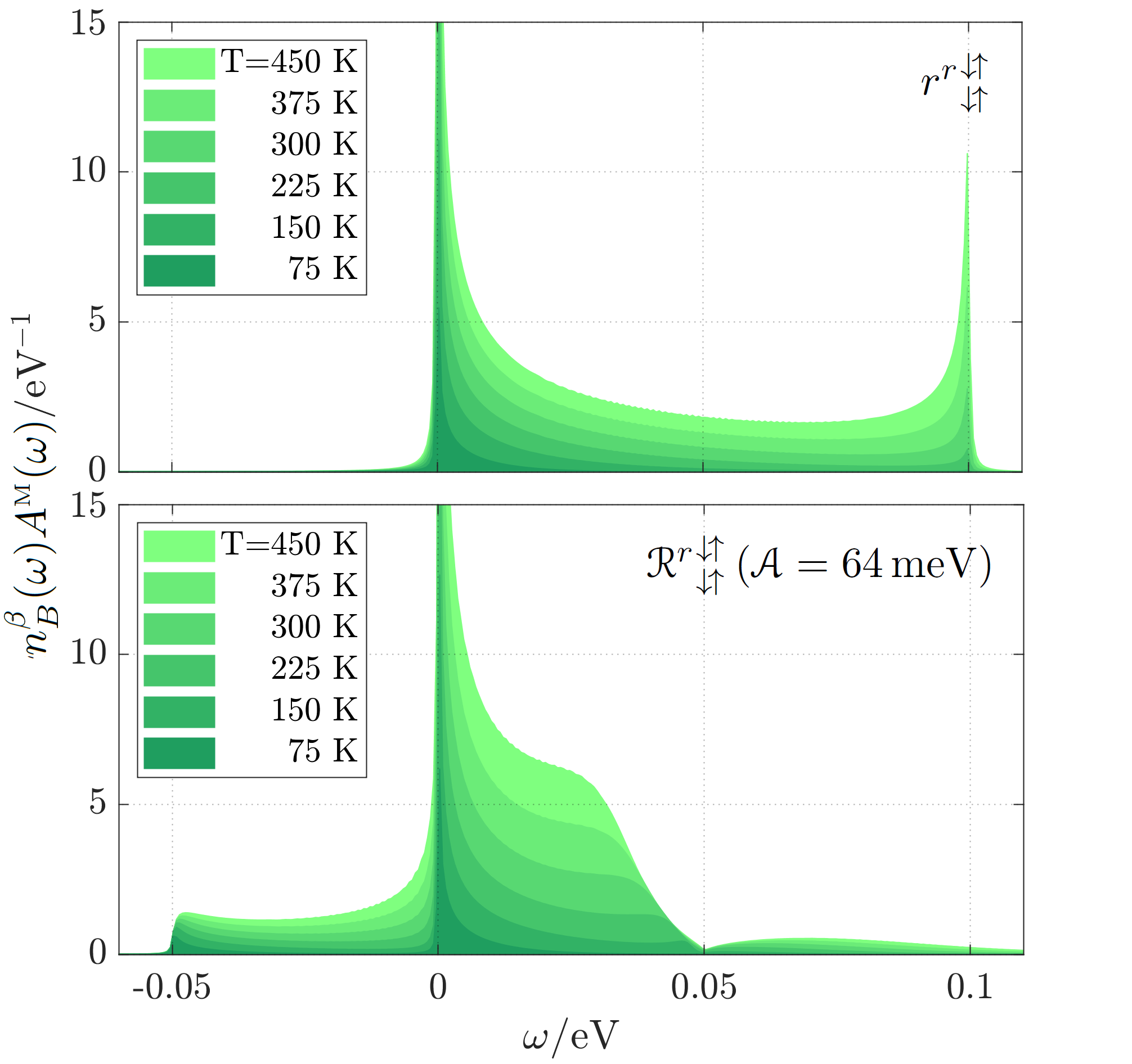}
\caption{Product between the Bose occupation ($n_B^\beta(\omega)$) and the total magnon spectrum, ($A^{\mathsmaller{\text{M}}}_{}(\omega)$), for Fock magnons (${r^r}^{\downarrow \! \uparrow}_{\downarrow \! \uparrow}$) and renormalized magnons (${\mathcal{R}^r}^{\downarrow \! \uparrow}_{\downarrow \! \uparrow}$) at various temperatures. $\mathcal{A}=64 \operatorname{meV}$.} 
\label{fig:fig5}
\end{figure} 
\begin{figure}[t] 
\includegraphics[width=0.95\linewidth]{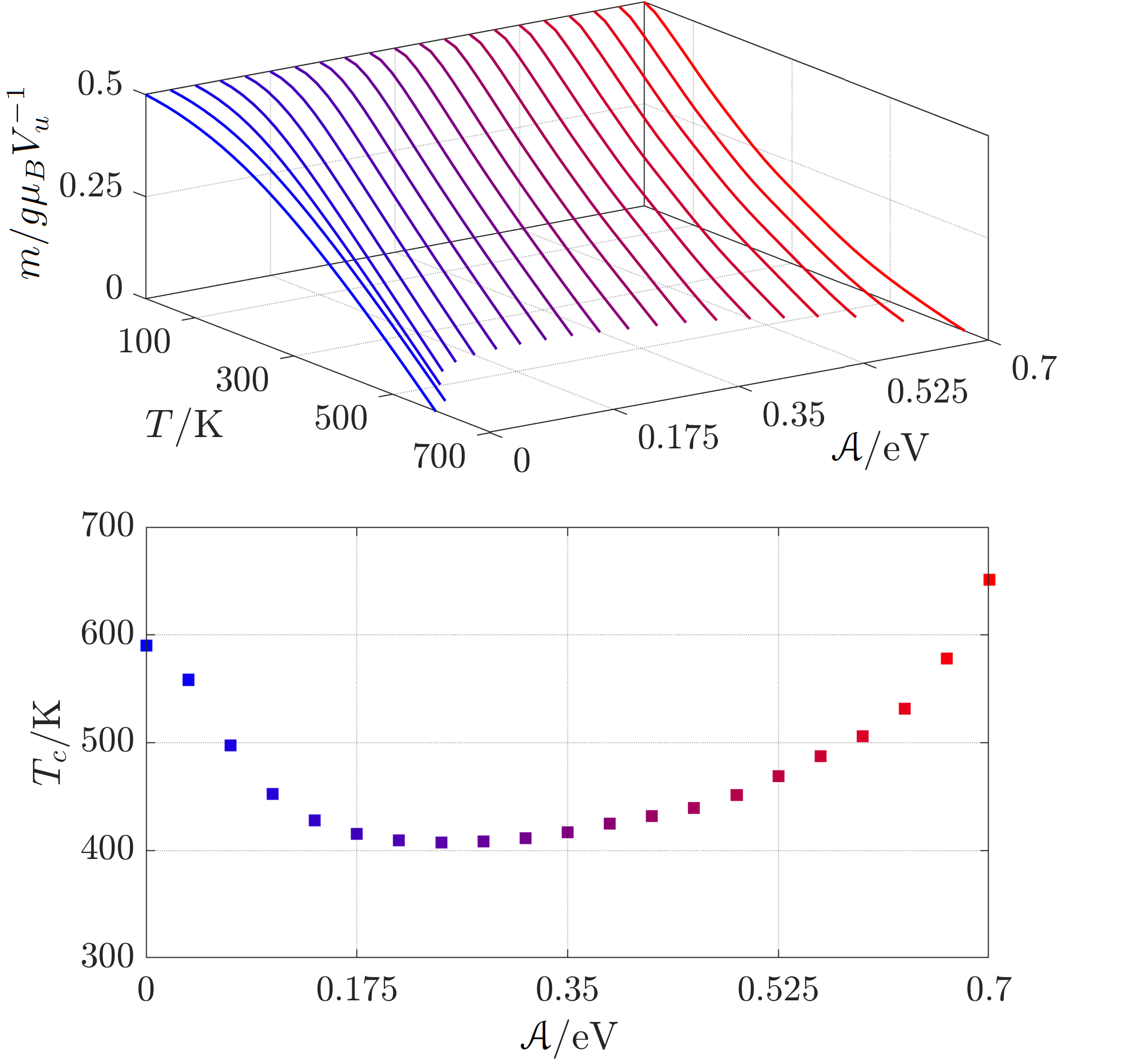}
\caption{Spin magnetization ($m$) as a function of $T$ for different values of $\mathcal{A}$, together with the associated Curie temperatures ($T_c$) for which $m=0$.} 
\label{fig:fig6}
\end{figure}

The magnitude of the renormalized magnon (RM) spectral function, $A^{\mathsmaller{\text{M}}}_{}(k,\omega) = -\tfrac{1}{\pi} \operatorname{Im} {\mathcal{R}^r}^{\uparrow \! \downarrow}_{\uparrow \! \downarrow}(k,\omega)$, is shown in Fig. \ref{fig:fig3} for different values of $\mathcal{A}$, at $T=0\operatorname{K}$ and $T=300\operatorname{K}$. With increased $\mathcal{A}$, the zero-temperature RM spectrum acquires an increasingly significant background continuum in the energy-range between $0.05\operatorname{eV}$ and $0.15\operatorname{eV}$, where the imaginary-part of ${\Delta_{\mathsmaller{\text{MP}}}^{r}}$ dominates, as well as spectral weight above this energy-range, where the real-part dominates, see Fig. \ref{fig:fig2}. At room temperature, the background continuum stretches between $-0.05\operatorname{eV}$ and $0.15\operatorname{eV}$, and in addition to a temperature-induced shift of the spectral weight above this energy-range a splitting of the RM band emerges around the energy of $0.05\operatorname{eV}$ --- the energy difference between the FM band maximum and the phonon energy. 

The magnitude of the total RM spectral function, $A^{\mathsmaller{\text{M}}}_{}(\omega)  = \int_0^K \frac{4\pi k^2 d k}{\Omega_{\mathsmaller{\text{BZ}}}^{} } A^{\mathsmaller{\text{M}}}_{}(k,\omega)  $, is presented in Fig. \ref{fig:fig4} for different values of $\mathcal{A}$ at absolute zero and room temperature, and compared with the results of the FM. The temperature-induced splitting at the energy $0.05 \operatorname{eV}$ is shown to increase the low-energy spectral weight. The product between the Bose occupation and the total spectrum, $n^\beta_B(\omega) A^{\mathsmaller{\text{M}}}_{}(\omega)$, which is strictly positive and proportional to the temperature-dependent probability of finding a magnon with energy $\omega$, is presented in Fig. \ref{fig:fig5} at different temperatures for the FM and the RM, using $\mathcal{A}=64\operatorname{meV}$. Since each magnon carries angular momentum $1$, the frequency-integral of this product precisely yields the temperature-induced reduction of the spin magnetization, after dividing by the unit cell volume $V_u$ and the product $g\mu_B=1$, where $g = 2$ is the non-relativistic g-factor and $\mu_B=\tfrac{1}{2}$ the Bohr magneton. Assuming that the spin magnetization per unit volume at $T=0$ is $0.5 g \mu_B$, the temperature-dependent magnetization is shown in Fig. \ref{fig:fig6} for different values of $\mathcal{A}$. The $\mathcal{A}$-dependent Curie temperature for which the magnetization vanishes is also presented in the same figure. For $\mathcal{A}=0$ and low temperatures the thermally accessible magnons can be approximated as parabolic so that the famous $T^{3/2}$ law for the magnetization is reproduced. When increasing $\mathcal{A}$ from 0, the Curie temperature decreases due to an increase in the low-energy RM spectral weight, originating from the band narrowing below $0.05\operatorname{eV}$ due to the splitting. However, when continuing to increase $\mathcal{A}$ the Curie temperature starts to increase again as a consequence of a washing out of the RM spectral weight. 
\section{Summary and Outlook}
In this work, we have presented a formalism tailored for describing the magnon number-conserving magnon-phonon coupling when the orbital magnetic moment is quenched and spin-orbit coupling can be neglected. 

By writing the Coulomb interaction in a crossing symmetric way and using Schwinger's functional derivative method, we identified the {\it screened collective} four-point interaction $\mathcal{W}$ as a natural quantity to do the many-body expansion in. After residing to a Hubbard-like model, we showed by an iterative process that the spin-flip interaction acquires terms quadratic in $\mathcal{W}$ that describe the simultaneous propagation of spin- and charge-fluctuations. We showed how this spin-flip interaction can be contracted to a two-point quantity, and were then, after relaxing the clamped-nuclei approximation, able to arrive at a magnon-phonon coupling (Eq. \eqref{najsestnew5}) of the same form as obtained using phenomenological magnetoelastic models. Since the inputs to the magnon-phonon coupling in Eq. \eqref{najsestnew5} are accessible from first principle, it may be considered an {\it ab initio} expression. It is also easily extended to the multiband and multi phonon-mode case. To test the formula, a minimal model was studied, with isotropic magnons and dispersion-free phonons. A temperature-induced low-energy splitting of the magnon spectrum was found, which reduced the Curie temperature for small non-zero magnon-phonon coupling strengths.  

The formalism developed in this work will be extended to account for spin-orbit coupling and other semi-relativistic effects in a coming publication. Another line of inquiry is to investigate, from first principles, how the magnon-phonon coupling in Eq. \eqref{najsestnew5} affects the critical temperature in high-temperature superconductors, by studying the superconducting gap equation.   
\begin{acknowledgments} We acknowledge valuable discussions with Silke Biermann and financial support from the Swedish Research Council (VR).
\end{acknowledgments}
  

\begin{thebibliography}{9}
\bibitem{liao2016}
B. Liao, A. A. Maznev, K. A. Nelson, and G. Chen, Nat. Commun. \textbf{7}, 13174 (2016). 
\bibitem{Eichenfield2009}
M. Eichenfield, J. Chan, R. M. Camacho, K. J. Vahala, and O. Painter, Nature \textbf{462}, 78–82 (2009).
\bibitem{aspel}
M. Aspelmeyer, T. J. Kippenberg, and F. Marquardt, Rev. Mod. Phys. \textbf{86}, 1391 (2014).
\bibitem{serga2010} 
A. A. Serga, A. V. Chumak, and B. Hillebrands, J. Phys. D: Appl. Phys. \textbf{43} 264002 (2010).
\bibitem{cornelissen2015}
L. J. Cornelissen, J. Liu, R. A. Duine, J. Ben Youssef, and B. J. van Wees, Nat. Phys. \textbf{11}, 1022–1026 (2015).
\bibitem{chumak2015}
A. V. Chumak, V. I. Vasyuchka, A. A. Serga, and B. Hillebrands, Nat. Phys. \textbf{11}, 453-461 (2015).
\bibitem{demokritov2006}
S. O. Demokritov, V. E. Demidov, O. Dzyapko, G. A. Melkov, A. A. Serga, B. Hillebrands, and A. N. Slavin, Nature \textbf{443}, 430–433 (2006).
\bibitem{ruckrieger2015}
A. Rückriegel, and P. Kopietz, Phys. Rev. Lett. \textbf{115}, 157203 (2015).
\bibitem{scalapino2012}
 D. J. Scalapino, Rev. Mod. Phys. \textbf{84}, 1383 (2012).
\bibitem{essenberger2014} 
F. Essenberger, A. Sanna, A. Linscheid, F. Tandetzky, G. Profeta, P. Cudazzo, and E. K. U. Gross, Phys. Rev. B \textbf{90}, 214504 (2014). 
\bibitem{dai2000}
P. Dai, H. Y. Hwang, J. Zhang, J. A. Fernandez-Baca, S.-W. Cheong, C. Kloc, Y. Tomioka, and Y. Tokura, Phys. Rev. B \textbf{61}, 9553 (2000). 
\bibitem{berk2019}
C. Berk, M. Jaris, W. Yang, S. Dhuey, S. Cabrini, and H. Schmidt, Nat. Commun. \textbf{10}, 2652 (2019) 
\bibitem{berk2020} 
C. R. Berk, and H. Schmidt, Physics \textbf{13}, 167 (2020).
\bibitem{man2017}
H. Man, Z. Shi, G. Xu, Y. Xu, X. Chen, S. Sullivan, J. Zhou, K. Xia, J. Shi, and P. Dai, Phys. Rev. B \textbf{96}, 100406(R) (2017).
\bibitem{ramachandran2015}
B. Ramachandran, K. K. Wu, Y. K. Kuo, and M. S. Ramachandra Rao,  J. Phys. D: Appl. Phys. \textbf{48} 115301 (2015).
\bibitem{aguilar2007}
R. Valdés Aguilar, A. B. Sushkov, C. L. Zhang, Y. J. Choi, S.-W. Cheong, and H. D. Drew, Phys. Rev. B \textbf{76}, 060404(R) (2007).
\bibitem{Kormann2014}
F. Körmann, B. Grabowski, B. Dutta, T. Hickel, L. Mauger, B. Fultz, and J. Neugebauer, Phys. Rev. Lett. \textbf{113}, 165503 (2014).
\bibitem{Perera2017}
D. Perera, D. M. Nicholson, M. Eisenbach, G. Malcolm Stocks, and David P. Landau, Phys. Rev. B \textbf{95}, 014431 (2017).
\bibitem{jaworski2011}
C. M. Jaworski, J. Yang, S. Mack, D. D. Awschalom, R. C. Myers, and J. P. Heremans, Phys. Rev. Lett. \textbf{106}, 186601 (2011).
\bibitem{kikkawa2016}
T. Kikkawa, K. Shen, B. Flebus, R. A. Duine, K. Uchida, Z. Qiu, G. E. W. Bauer, and E. Saitoh, Phys. Rev. Lett. \textbf{117}, 207203 (2016).
\bibitem{flebus2017}
B. Flebus, K. Shen, T. Kikkawa, K. Uchida, Z. Qiu, E. Saitoh, R. A. Duine, and G. E. W. Bauer, Phys. Rev. B \textbf{95}, 144420 (2017).
\bibitem{yahiro2020}
R. Yahiro, T. Kikkawa, R. Ramos, K. Oyanagi, T. Hioki, S. Daimon, and E. Saitoh, Phys. Rev. B
\textbf{101}, 024407 (2020).
\bibitem{uchida2011}
K. Uchida, H. Adachi, T. An, T. Ota, M. Toda, B. Hillebrands, S. Maekawa, and E. Saitoh, Nat. Mater. \textbf{10}, 737–741 (2011).
\bibitem{bauer2012}
G. E. W. Bauer, E. Saitoh, and B. J. van Wees, Nat. Mater. \textbf{11}, 391–399 (2012). 
\bibitem{weiler2012}
M. Weiler, H. Huebl, F. S. Goerg, F. D. Czeschka, R. Gross, and S. T. B. Goennenwein, Phys. Rev. Lett. \textbf{108}, 176601 (2012).
\bibitem{hayashi2018}
H. Hayashi, and K. Ando, Phys. Rev. Lett. \textbf{121}, 237202 (2018). 
\bibitem{sharma2004}
P. A. Sharma, J. S. Ahn, N. Hur, S. Park, S. B. Kim, S. Lee, J.-G. Park, S. Guha, S.-W. Cheong, Phys. Rev. Lett. \textbf{93}, 177202 (2004).
\bibitem{katsura2010}
H. Katsura, N. Nagaosa, and P. A. Lee, Phys. Rev. Lett. \textbf{104}, 066403 (2010).
\bibitem{zhang2014}
L. Zhang, and Q. Niu, Phys. Rev. Lett. \textbf{112}, 085503 (2014).
\bibitem{Garanin2015}
D. A. Garanin, and E. M. Chudnovsky, Phys. Rev. B \textbf{92}, 024421 (2015).
\bibitem{holanda2018}
J. Holanda, D. S. Maior, A. Azevedo, and S. M. Rezende, Nat. Phys. \textbf{14}, 500–506 (2018).
\bibitem{ogawa2015}
N. Ogawa, W. Koshibae, A. J. Beekman, N. Nagaosa, M. Kubota, M. Kawasaki, and Y. Tokura, Proc. Natl. Acad. Sci. U.S.A. \textbf{112}, 8977 (2015).
\bibitem{zhang2019}
X. Zhang, Y. Zhang, S. Okamoto, and D. Xiao, Phys. Rev. Lett. \textbf{123}, 167202 (2019).
\bibitem{an2016}
K. An, K. S. Olsson, A. Weathers, S. Sullivan, X. Chen, X. Li, L. G. Marshall, X. Ma, N. Klimovich, J. Zhou, L. Shi, and X. Li, Phys. Rev. Lett. \textbf{117}, 107202 (2016).
\bibitem{struzhkin2016}
V. V. Struzhkin, Low Temp. Phys. \textbf{42}, 884 (2016).
\bibitem{bozhko2017}
D. A. Bozhko, P. Clausen, G. A. Melkov, V. S. L’vov, A. Pomyalov, V. I. Vasyuchka, A. V. Chumak, B. Hillebrands, and A. A. Serga, Phys. Rev. Lett. \textbf{118}, 237201 (2017).
\bibitem{lifshitz1935}
L. D. Landau, and E. M. Lifshitz, Phys. Z. Sowjetunion \textbf{8}, 101 (1935).
\bibitem{abrahams1952}
E. Abrahams, and C. Kittel, Phys. Rev. \textbf{88}, 5 (1952). 
\bibitem{abrahams1954}
C. Kittel, and E. Abrahams, Phys. Rev. \textbf{25}, 1 (1953). 
\bibitem{kittel1958}
C. Kittel, Phys. Rev. \textbf{110}, 836 (1958). 
\bibitem{kaganov1959}
M. I. Kaganov, and V. M. Tsukernik, Sov. Phys. JETP \textbf{9}, 151
(1959).
\bibitem{tiersten1964}
H. F. Tiersten, J. Math. Phys. \textbf{5}, 1298 (1964).
\bibitem{schlomann1964}
E. Schlömann, and R. Joseph, J. Appl. Phys. \textbf{35}, 2382 (1964).
\bibitem{pytte1965}
E. Pytte, Ann. Phys. (NY) \textbf{32}, 377 (1965).
\bibitem{silberglitt1969}
R. Silberglitt, Phys. Rev. \textbf{188}, 786 (1969).
\bibitem{rezende1969}
S. M. Rezende, and F. R. Morgenthaler, J. Appl. Phys. \textbf{40}, 524 (1969).
\bibitem{guerreiro1971}
S. C. Guerreiro, and S. M. Rezende, Rev. Bras. Fis. \textbf{1}, 207 (1971).
\bibitem{jensen1975}
J. Jensen, and J. G. Houmann, Phys. Rev. B \textbf{12}, 320 (1975). 
\bibitem{economou1975}
E. N. Economou, K. L. Ngai, T. L. Reinecke, J. Ruvalds, and Richard Silberglitt, Phys. Rev. B \textbf{13}, 3135 (1976).
\bibitem{sanger1994}
D. U. Sänger, Phys. Rev. B \textbf{49}, 12176 (1994).
\bibitem{woods2001}
L. M. Woods, Phys. Rev. B \textbf{65}, 014409 (2001).
\bibitem{cheng2007}
T.-M. Cheng and L. Li, J. Magn. Magn. Mater. \textbf{320}, 1 (2008).
\bibitem{berciu2009}
M. Berciu, and G. A. Sawatzky, Phys. Rev. B \textbf{79}, 195116 (2009).
\bibitem{ruckriegel2014}
A. Rückriegel, P. Kopietz, D. A. Bozhko, A. A. Serga, and B. Hillebrands, Phys. Rev. B \textbf{89}, 184413 (2014).
\bibitem{Guerreiro2015}
S. C. Guerreiro, and S. M. Rezende, Phys. Rev. B \textbf{92}, 214437 (2015).
\bibitem{kamra2015}
A. Kamra, H. Keshtgar, P. Yan, and G. E. W. Bauer, Phys. Rev. B, \textbf{91}, 104409 (2015).
\bibitem{chernyshev2015}
A. L. Chernyshev, and W. Brenig, Phys. Rev. B \textbf{92}, 054409 (2015).
\bibitem{takahashi2016}
R. Takahashi, and N. Nagaosa, Phys. Rev. Lett. \textbf{117}, 217205 (2016).
\bibitem{Maehrlein2018}
S. F. Maehrlein, I. Radu, P. Maldonado, A. Paarmann, M. Gensch, A. M. Kalashnikova, R. V. Pisarev, M. Wolf, P. M. Oppeneer, J. Barker, and T. Kampfrath, Sci. Adv. \textbf{4}, eaar5164 (2018).
\bibitem{schmidt2018}
R. Schmidt, F. Wilken, T. S. Nunner, and P. W. Brouwer, Phys. Rev. B \textbf{98}, 134421 (2018).
\bibitem{holm2018}
S. L. Holm, A. Kreisel, T. K. Schäffer, A. Bakke, M. Bertelsen, U. B. Hansen, M. Retuerto, J. Larsen, D. Prabhakaran, P. P. Deen, Z. Yamani, J. O. Birk, U. Stuhr, Ch. Niedermayer, A. L. Fennell, B. M. Andersen, and K. Lefmann, Phys. Rev. B \textbf{97}, 134304 (2018).
\bibitem{rameshti2019}
B. Z. Rameshti, and R. A. Duine, Phys. Rev. B \textbf{99}, 060402(R) (2019).
\bibitem{Streib2019}
S. Streib, N. Vidal-Silva, K. Shen, and G. E. W. Bauer, Phys. Rev. \textbf{B} 99, 184442 (2019). 
\bibitem{antropov1996}
V. P. Antropov, M. I. Katsnelson, B. N. Harmon, M. van Schilfgaarde, and D. Kusnezov, Phys. Rev. B \textbf{54}, 1019 (1996).
\bibitem{Ciornei2011}
M.-C. Ciornei, J. M. Rubí, and J.-E. Wegrowe
Phys. Rev. B \textbf{83}, 020410(R) (2011).
\bibitem{Bhattacharjee2012}
Satadeep Bhattacharjee, Lars Nordström, and Jonas Fransson
Phys. Rev. Lett. \textbf{108}, 057204 (2012).
\bibitem{Giustino2017}
Rev. Mod. Phys. \textbf{89}, 015003 (2017).
\bibitem{hedin69}
L. Hedin and S. Lundqvist, in {\it Solid State Physics, Advanced
in Research and Applications}, edited by F. Seitz, D. Turnbull, and H. Ehrenreich (Academic, New York, 1969), Vol. 23, pp. 1–181.
\bibitem{fierz} 
T Ayral, J Vučičević, and Olivier Parcollet, Phys. Rev. Lett. \textbf{119}, 166401 (2017).
\bibitem{firsthedin}
L. Hedin, Phys. Rev. \textbf{139}, A796 (1965).
\bibitem{starke}
R. Starke, and G. A. H. Schober, Phot. Nano. Fund. Appl. \textbf{14}, 1
(2015).
\bibitem{reining2002}
G. Onida, L. Reining, and A. Rubio, Rev. Mod. Phys. \textbf{74}, 601 (2002).
\bibitem{kutepov2017}
A. L. Kutepov, and G. Kotliar, Phys. Rev. B \textbf{96}, 035108 (2017).
\bibitem{scrT}
E. Sasioglu, A. Schindlmayr, C. Friedrich, F. Freimuth, and S. Blügel, Phys. Rev. B \textbf{81}, 054434 (2010).
\bibitem{Jensen}
J. Jensen, and G. Houmann, Phys. Rev. B \textbf{12}, 320 (1974).
\bibitem{Cheng}
T.-M. Cheng, and L. Li, J. Magn. Magn. Mater. \textbf{320} 1 (2008).
\bibitem{Hellsvik}
J. Hellsvik, D. Thonig, K. Modin, D. Iuşan, A. Bergman, O. Eriksson, L. Bergqvist, and A. Delin, Phys. Rev. B \textbf{99}, 104302 (2019).
\bibitem{jacob}
C. R. Jacob and M. Reiher, Int. J. Quantum Chem. \textbf{112}, 3661 (2012).
\bibitem{Goodvin2008}
G. L. Goodvin, and M. Berciu, Phys. Rev. B \textbf{78}, 235120 (2008).
\end{thebibliography}
\end{document}